\documentclass[12pt,draftclsnofoot,onecolumn,a4]{IEEEtran}
\usepackage[utf8]{inputenc}
\usepackage{url}

\usepackage{cite}
\usepackage{graphicx}
\usepackage{booktabs}
\usepackage{array}
\usepackage{dcolumn}
\usepackage{blindtext}
\usepackage{scrextend}
\usepackage{array}
\usepackage{arydshln}
\usepackage{amsmath,amssymb}
\usepackage{booktabs}
\usepackage{multirow}
\usepackage{algorithm,algorithmic}
\usepackage{amsmath}
\usepackage{array}
\usepackage{xcolor}
\usepackage{setspace}
\usepackage{amsthm}
\usepackage {hyperref}

\usepackage{titlesec}
 \usepackage{url}
\titlespacing{\section}{0pt}{2pt}{0pt}
\allowdisplaybreaks
\usepackage{bbm}
\usepackage[mathscr]{euscript}
 \let\mathscr\relax
\usepackage[scr]{rsfso}

\newtheorem{proposition}{{\it Proposition}}

\makeatletter
\g@addto@macro\normalsize{%
 \setlength\abovedisplayskip{2pt}
 \setlength\belowdisplayskip{2pt}
 \setlength\abovedisplayshortskip{2pt}
 \setlength\belowdisplayshortskip{2pt}
}
\usepackage{titlesec}
\titlespacing{\section}{0pt}{2pt}{0pt}

\newcommand{\sfur}{\mathsf{ur}}
\newcommand{\sfem}{\mathsf{em}}
\newcommand{\pro}{\mathsf{pro}}
\newcommand{\tx}{\mathsf{tx}}

\usepackage{enumitem}
\usepackage{tikz}
\newcommand*\circled[1]{\tikz[baseline=(char.base)]{%
            \node[shape=circle,fill=white!20,draw,inner sep=2pt] (char) {#1};}}

\title{\huge Intelligent Traffic Steering in Beyond 5G Open RAN based on LSTM Traffic Prediction}
\author{\IEEEauthorblockN{Fatemeh Kavehmadavani, Van-Dinh Nguyen, \\ Thang X. Vu, and Symeon Chatzinotas}\\

\thanks{The authors are with the Interdisciplinary Centre for Security, Reliability
and Trust (SnT), University of Luxembourg (email: \{fatemeh.kavehmadavani, dinh.nguyen, thang.vu, symeon.chatzinotas\}@uni.lu). This work was supported in part by the ERC AGNOSTIC project, ref. H2020/ERC2020POC/957570/DREAM, and by the FNR CORE ASWELL project, ref. FNR/C19/IS/13718904/ASWELL/Chatzinotas.}
}

\begin{document}

\maketitle

\begin{abstract}
Open radio access network (ORAN) Alliance offers a disaggregated RAN functionality built using open interface specifications between blocks. To efficiently support various competing services, \textit{namely} enhanced mobile broadband (eMBB) and ultra-reliable and low-latency (uRLLC), the ORAN Alliance has introduced a standard approach toward more virtualized, open and intelligent networks. To realize benefits of ORAN in optimizing resource utilization, this paper studies an intelligent traffic steering (TS) scheme within the proposed disaggregated ORAN architecture. For this purpose, we propose a joint intelligent traffic prediction, flow-split distribution, dynamic user association and radio resource management (JIFDR) framework in the presence of unknown dynamic traffic demands. To adapt to dynamic environments on different time scales, we decompose the formulated optimization problem into two long-term and short-term subproblems, where the optimality of the later is strongly dependent on the optimal dynamic traffic demand. We then apply a long-short-term memory (LSTM) model to effectively solve the long-term subproblem, aiming to predict dynamic traffic demands, RAN slicing, and flow-split decisions. The resulting non-convex short-term subproblem is converted to a more computationally tractable form by exploiting successive convex approximations. Finally, simulation results are provided to demonstrate the effectiveness of the proposed algorithms compared to several well-known benchmark schemes. 
\end{abstract}

\begin{IEEEkeywords}
Beyond 5G networks, open radio access networks, intelligent resource management, traffic prediction, traffic steering, long short-term memory,  network slicing.
\end{IEEEkeywords}

\newpage
\section{Introduction}\label{section1}
Next-generation (``NextG") mobile communication networks (\textit{e.g.}, beyond fifth-generation (5G) and sixth-generation (6G)) are designed to accommodate a wide range of service types with their own specific demands, such as throughput, reliability and delay. The mentioned services are basically categorized into three principal cases, enhanced mobile broadband (eMBB), massive machine-type communications (mMTC) and ultra-reliability low-latency communication (uRLLC) \cite{popovski20185g}. Efficiently supporting the coexistence of these heterogeneous services is challenging in the ``NextG" wireless networks due to their competing demands. The existing ``one-size-fits-all" 5G architecture makes it very difficult if not impossible to enable the coexistence of heterogeneous services since the present 5G wireless networks are aggregated, closed and inflexible. Despite the cost-effectiveness of centralized/cloud radio access networks (CRAN) and virtual radio access networks (vRAN), open interfaces, non-proprietary hardware and software are still lacking in these systems. Open RAN (ORAN) is an emerging solution to enable flexible, virtualized, disaggregated, intelligent and open ``NextG" wireless networks to support heterogeneity of wireless services \cite{gavrilovska2020cloud}. The openness of RAN components not only increases the interoperability between vendors but also speeds up the delivery of new services, which can be dynamically nominated to users. Due to the increasing complexity of ``NextG" wireless networks, a self-organizing network's optimization, deployment and operation are increasingly becoming impossible without intelligence  \cite{wang2020artificial}, \cite{niknam2020federated}. 

Accommodating heterogeneous services (uRLLc, eMBB and mMTC) with competing demands on the identical RAN infrastructure is exceedingly challenging, such that building numerous physical networks to accommodate distinct services is not practical. Hence, it is difficult to efficiently route heterogeneous traffics to enhance user experience and network efficiency \cite{dryjanski2016unified}. To this end, the concept of RAN slicing has been suggested as a potential remedy to constantly assign the accessible storage, compute and communication resources across multiple services whilst guaranteeing their isolation \cite{vassilaras2017algorithmic}. In addition, traffic steering (TS) is considered as one of the most efficient approaches that enables network software to steer the traffics in the most proper paths. Nevertheless, the available research on TS in 5G is still limited and uncompleted. For example, the typical TS treats all users similarly, regardless of users' demands and network conditions, meaning that a network operator may even be wasted its resources if a simple strategy is implemented. 
To enhance throughput and reliability in wireless networks with limited bandwidth, the multi-connectivity (MC) technique can be used to aggregate multiple links and allow a user to connect to more than two nodes. In practice, MC has the potential to dramatically reduce interference and latency of mobility methods, especially at the cell edge \cite{suer2019multi}. The multi-link capability makes MC the most practical method for achieving uRLLC and eMBB coexistence, whereas the recent proposals for the 5G air interface in 3GPP Release 15 utilize flexible mixed numerologies \cite{arslan2018flexible}.

Since embedding intelligence in ORAN is a forward factor, this paper introduces a joint intelligent traffic prediction and radio resource management framework, taking into account the ORAN architectural requirements and various service requirements. This paper benefits from the long short-term memory (LSTM) recurrent neural network (RNN) to learn the network traffic pattern and predict the incoming traffic packets of the network. LSTM has been introduced as an undeniable state-of-the-art method within the deep neural networks to overcome the exploding/vanishing gradient problem, especially in learning long-term dependencies \cite{yu2019review}. We outline the compliance of the overall scheme with the ORAN requirements later.

\subsection{Related Works}
To improve services for network providers, the work in \cite{kamel2014lte} focused on providing an efficient scheduling scheme to dynamically allocate radio resources in LTE networks. In \cite{pocovi2018joint}, the authors proposed a joint resource allocation and dynamic link adaptation scheme for multiplexing eMBB and uRLLC on a shared channel, which dynamically tunes the block error probability of URLLC small payload transmissions in each cell. A control channel and packet size aware resource allocation approach was introduced in \cite{karimi2019efficient} to enable the packet scheduling and resource allocation for uRLLC and eMBB traffics coexistence in 5G NR networks.
Although the heuristic algorithm proposed in \cite{karimi2019efficient} meets the uRLLC's requirements by preserving a large number of resources to uRLLC, this method has failed to isolate the slice, resulting in a reduction of the eMBB throughput compared to high uRLLC traffic. Wu \textit{et al.}  \cite{wu2017signal} developed the puncturing method to eliminate the uRLLC queuing delay for multiplexing of uRLLC and eMBB services. The authors in \cite{anand2020joint} studied a joint scheduling scheme to maximize the eMBB throughput while minimizing the utility of uRLLC to meet the quality of service (QoS) requirements. Since uRLLC services are prioritized in the puncturing-based schemes and scheduled on the assigned eMBB's resources, the eMBB performance (throughput and reliability) significantly decreases when the uRLLC traffic increases. Moreover, the fixed-numerology over frequency-time resources for the scheduling scheme is often considered.

There is significant attention from academia and industry about TS in the literature. In \cite{zhang2016beyond}, a TS framework was studied in unlicensed bands on the LTE network in order to distribute traffic among radio access technologies, heterogeneous cells and spectrum bands. To overcome the puncturing difficulties in multiple services, Praveenkumar \textit{et al.} in \cite{korrai2020ran} proposed a slice-isolated RAN slicing scheme with orthogonal frequency-division multiple access (OFDMA) for the coexistence of uRLLC and eMBB. A joint scheduling and TS scheme based on dynamic MC and RAN slicing in 5G networks were analyzed in \cite{zhang2020dynamic}, in which an effective capacity model to evaluate the frameworks' performance is proposed. To integrate the LTE into 5G networks, Prasad\textit{et al.} \cite{prasad2016enabling} investigated an energy-efficient RAN moderation and dynamic TS  based on the connectivity by multiple radio links.

The RAN slicing framework over multiple services networks has been recently developed under frequency-time resources thanks to the flexibility of mixed-numerologies. The authors in \cite{you2018resource} studied a resource allocation optimization problem by considering the flexible numerology in both frequency and time domains. The work in \cite{nguyen2019wireless} analyzed the wireless scheduling optimization problem over the mixed-numerologies to support the heterogeneous services with different QoS requirements, assuming that mapping the radio resources (time and frequency) is decoupled from service scheduling. A joint optimization of RAN slicing, resource block and power allocation problem for eMBB, mMTC and uRLLC in 5G wireless networks was considered in \cite{korrai2020joint} under imperfect channel state information (CSI).

However, the aforementioned works have investigated TS with flexible numerology in the ``one-size-fits-all" network architecture, which is not adaptable enough to support heterogeneous services. Despite the huge benefit of intelligence of ORAN, there are only a few attempts on the TS in the literature. Solmaz \textit{et al.} in \cite{niknam2020intelligent} proposed an intelligent traffic prediction and radio resource management framework to control the congested cell based on cell-splitting in ORAN architecture for multiplexing uRLLC and eMBB services. In \cite{bonati2021intelligence}, a systematic analysis for implementing the intelligence in each layer of ORAN architecture for data-driven ``NexG" wireless networks was provided by considering the closed-control loops between ORAN components. Furthermore, in our previous study \cite{kavehmadavani2022traffic}, we have proposed a TS scheme based on MC and RAN slicing technologies to effectively allocate diverse network resources in ORAN architecture by assuming fixed-numerology (\textit{i.e.}, 0.25ms mini-slots) tailored with 5G NR. 

\subsection{Contributions}
In this paper, we develop an intelligent TS framework in the presence of unknown dynamic traffic demand to meet the requirements of both uRLLC and eMBB services in beyond 5G based on dynamic MC. Learning an optimal traffic steering policy in dynamic environments is challenging because fluctuations in traffic demand over time are non-stationary and unknown, hindering the computation of cost-efficient associations. This proposed framework is handled by rAPPs and xAPP at non-real-time RAN intelligent controller (non-RT RIC) and near-real-time RIC (near-RT RIC) of the ORAN architecture. The existing rAPPs at non-RT RIC include the traffic prediction, dynamic RAN slicing decision and flow-split distribution, while the xAPP at near-RT RIC is radio resource management to schedule the joint resource block and transmission power with mixed numerologies based on standardization in 5G NR. To the best of our knowledge, this is the first work to model intelligent TS in the ORAN architecture considering the mixed-numerology in the presence of unknown traffic demands.

To achieve the maximum throughput for eMBB traffic while guaranteeing the minimum uRLLC latency requirement and vice versa, we propose a joint intelligent traffic prediction, flow-split distribution, dynamic user association and radio resource management scheme befitting the ORAN architecture. Then, we identify the location of the ML training, AI server and inference modules to provide a high-level architecture of deployment scenarios and end-to-end flow to prove compatibility with ORAN standards. Our main contributions are summarized as follows:
\begin{itemize}
    \item We develop a general optimization framework to jointly optimize the intelligent traffic prediction, flow-split distribution, dynamic user association and radio resource management, called ``JIFDR''. To maximize the eMBB's throughput while guaranteeing the uRLLC latency requirement, or vice versa, we formulate two optimization problems with different objective designs while satisfying QoS requirements, slice isolation, power budget and maximum fronthaul (FH) capacity.
    \item To effectively solve the formulated problems, we divide each problem into the long-term and short-term subproblems, which are executed on different time scales. The long-term subproblem is mapped into three dependent rAPPs: traffic prediction, dynamic RAN slicing decision and flow-split distribution at the non-RT RIC. In contrast, the short-term sub-problem is deployed as the radio resource management xAPP at the near-RT RIC, which is linked to the upper layer through the A1 interface.
    \item The long-term subproblem benefits from the LSTM RNN to learn and predict traffic patterns and demands. This model is trained offline at the non-RT RIC in the service management and orchestration (SMO) through the long-term collected data from the RAN layer via the O1 interface. RNN is utilized to learn the temporal pattern of the traffic demand from current values in order to forecast future values. Upon the inference result, two heuristic methods are proposed to optimize the RAN slicing and flow-split distribution. 
    \item Next, given rAPPs' outcomes sent from the non-RT RIC via the A1 interface, we propose a successive convex approximation (SCA)-based iterative algorithm to solve the short-term subproblem, which belongs to a class of mixed-integer non-convex programming (MINCP) problem. 
    \item Finally, numerical results are presented to demonstrate the proposed algorithm's quick convergence behaviour and to confirm its efficacy in comparison to benchmark schemes. Furthermore, by using a mathematical analysis a convergence and complexity analysis is studied. The average mean square error (MSE) of the prediction is relatively low at $0.0033$.
\end{itemize}

The rest of this paper is organized as follows. Section \ref{section2} introduces the ORAN architecture and system model. In Section \ref{section3}, we present the problem formulation and overall intelligent TS deployment architecture and algorithm. Section \ref{section4} first proposes the LSTM model and heuristic methods to solve the long-term subproblem and then develops an SCA-based iterative algorithm to solve the short-term subproblem. Simulation results and discussions are provided in Section \ref{section6}, while Section \ref{section7} concludes the paper.

\section{ORAN Architecture and System Model} \label{section2}

\subsection{ORAN Architecture}
\begin{figure}[h]
\centering
    \includegraphics[width=0.6\textwidth,trim=2 2 2 2,clip=true]{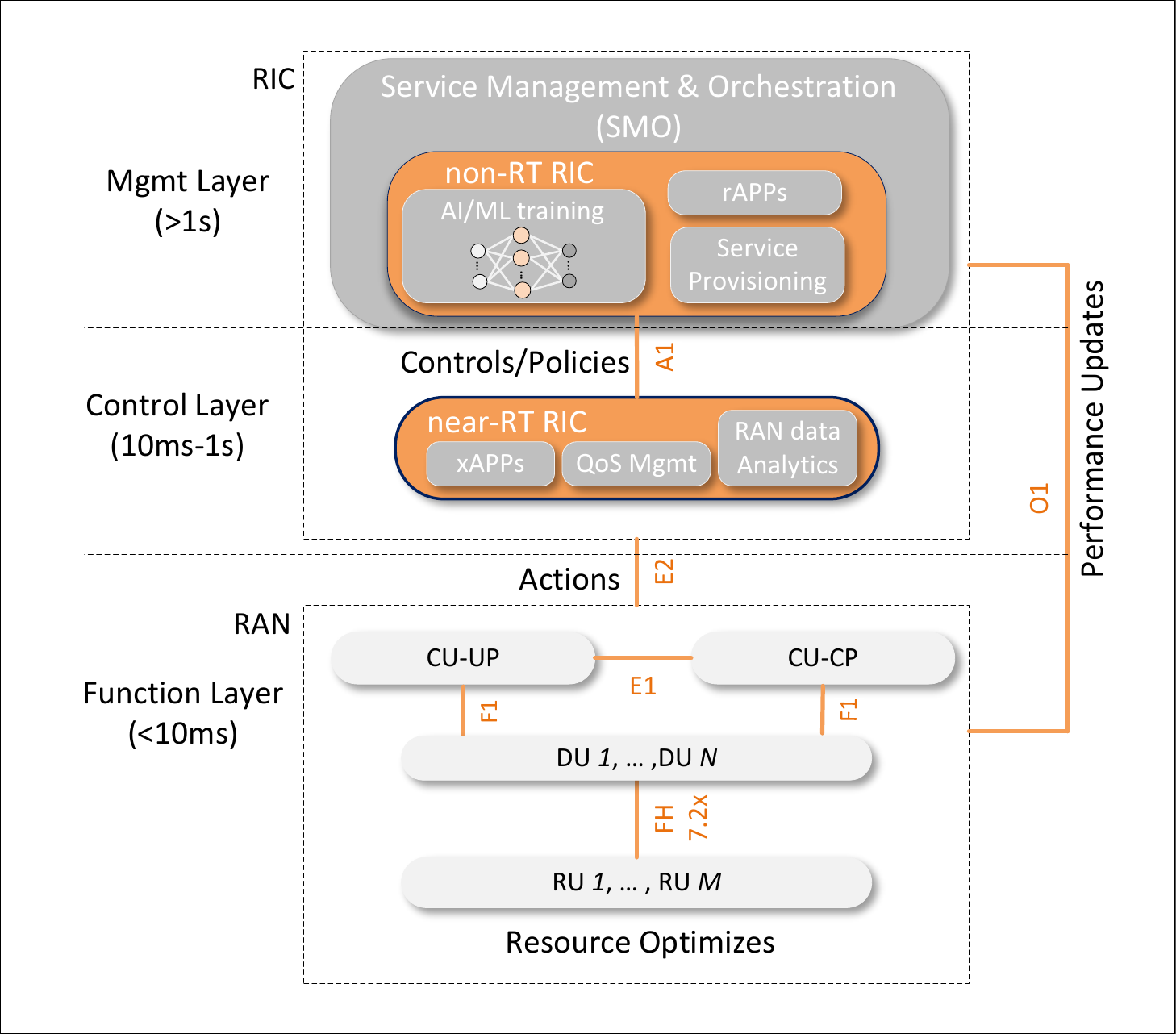}
    \caption{ORAN architecture based on ORAN Alliance \cite{oran2018}.}
    \label{fig1}
\end{figure}

The ORAN architecture based on the ORAN Alliance is illustrated in Fig. \ref{fig1}, including three main layers (the management, control and function layers). To further reduce the RAN expenditure, ORAN fosters self-organizing networks by adding two unique modules of near-RT and non-RT RICs to enable a centralized network abstraction which improves efficiency by cost-reducing the human-machine interaction. Following the disaggregation concept, BS functionalities are virtualized as network functions based on the 3GPP functional split and are distributed among various network nodes, \textit{namely} central unit (CU), distributed unit (DU) and radio unit (RU) \cite{niknam2020intelligent}. Hence, open interfaces (FH, A1, O1, E2, F1) are introduced to enable efficient multi-vendor interoperability, where a network operator can select RAN components from different vendors individually. 
 
 The unique feature of RICs is to create closed-control loops (\textit{i.e.}, autonomous action and feedback loops) between RAN components and their controllers. In order to control traffic prediction, network slicing and hand-over management, ORAN defines three control loops running at timescales ranging from $1$ ms to thousands of ms, enabling real-time control of transmission methods and beamforming. In particular, the non-RT RIC carries out tasks with a temporal granularity greater than one second, like service provisioning and training AI/ML models. On the other hand, the near-RT RIC manages operations with timescales of more than $10$ ms, hosts external applications (referred to as xApps) and incorporates intelligence in the RAN by data-driven control loops. Indeed, xAPPs are external applications specific to radio functions to make the RAN components programmable. To this end, ORAN Alliance strives to steer the industry towards the development of AI/ML-enabled RICs.

\subsection{Network Model}
\begin{figure}
\centering
    \includegraphics[width=0.8\textwidth,trim=5 5 5 5,clip=true]{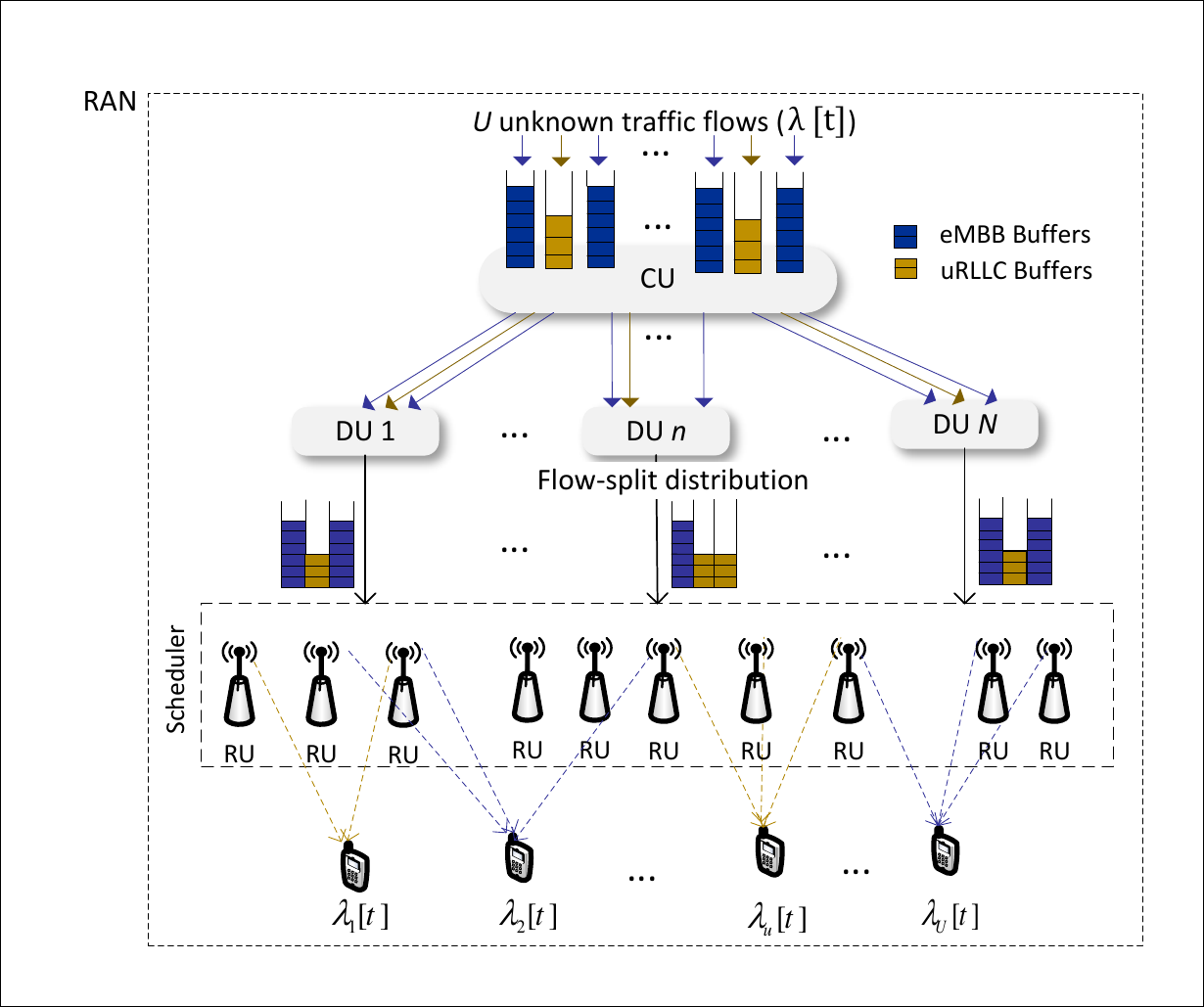}
    \caption{System model with the traffic-steering scheme.}
    \label{fig2}
\end{figure}
We consider a downlink OFDMA multi-user multiple-input single-output (MU-MISO) system in the ORAN architecture, consisting of one CU, the set $\mathscr{N}\triangleq\{1,2,\dots, N\}$ of $N$ DUs and the set $\mathscr{M} \triangleq\{1, 2, \dots, M\}$ of $M$ RUs. For cost-effective deployment, each DU serves a cluster of RUs. Let denote by $\mathscr{M}_n \triangleq\{(n,1), \dots ,(n,M_n)\}$ with $|\mathscr{M}_n|= M_n$ and $\sum_{n \in \mathscr{N}} M_n = M$ the set of RUs served by DU $n$. The $m$-th RU served by $n$-th DU is referred to as RU$(n,m)$, which is equipped with $K$ antennas while users are equipped with a single antenna.
Let us denote by $\mathscr{U} \triangleq \{1, \dots, U\}$ the set of users served by DUs, which can be further divided into two disjoint sets $\mathscr{U}^{\sfur}$ of $U^{\sfur}$ uRLLC users and $\mathscr{U}^{\sfem}$ of $U^{\sfem}$ eMBB users. The eMBB users generate the traffic with a large packet of size $Z^{\sfem}$ bytes, while uRLLC users generate a sequence of small and identical packets of $Z^{\sfur}$ bytes. 
In addition, as shown in Fig. {\ref{fig2}}, we assume that all data arriving from upper layers are stored in the user-specific transmission buffers of the RUs till it is time to serve it. The RUs serve the users in the cell by allocating the frequency-time radio resource blocks (RBs) and transmission power to each RB.

To meet the demands of exigent latency services, we investigate a mini-slot-based framework, where each time slot is broken into two mini-slots. Each mini-slot has a duration of $\delta= 1/ 2^{\gamma+1}$ ms and comprises $7$ OFDM symbols, where $\gamma \in \{0, 1, 2\}$ is the subcarrier spacing (SCS) index. Hereon, we suppose that several RUs operating in MC configuration are simultaneously providing eMBB and uRLLC services. Based on \cite{kihero2019inter}, numerology with index $i = 1$ (\textit{i.e.}, SCS index $\gamma = 1$) is appropriate for eMBB, which meets higher data rates, while due to the small data packet of uRLLC, numerology with index $i = 2$ (\textit{i.e.}, SCS index $\gamma = 2$) is more suitable for the uRLLC service's applications with latency-critical. From the mixed-numerologies point of view, eMBB service sorts the numerology $i=1$ with RB's bandwidth (BW) of $\beta_i|_{i=1}= 360$ kHz and $\delta_i|_{i=1}=0.25$ ms of transmission time interval (TTI) duration as the highest priority and uRLLC service would prioritize numerology $i=2$ with RB's BW of $\beta_i|_{i=2} =720$ kHz and $\delta_i|_{i=2}=0.125$ ms of TTI duration.

\begin{figure}
\centering
    \includegraphics[width=0.95\textwidth,trim=2 2 2 2,clip=true]{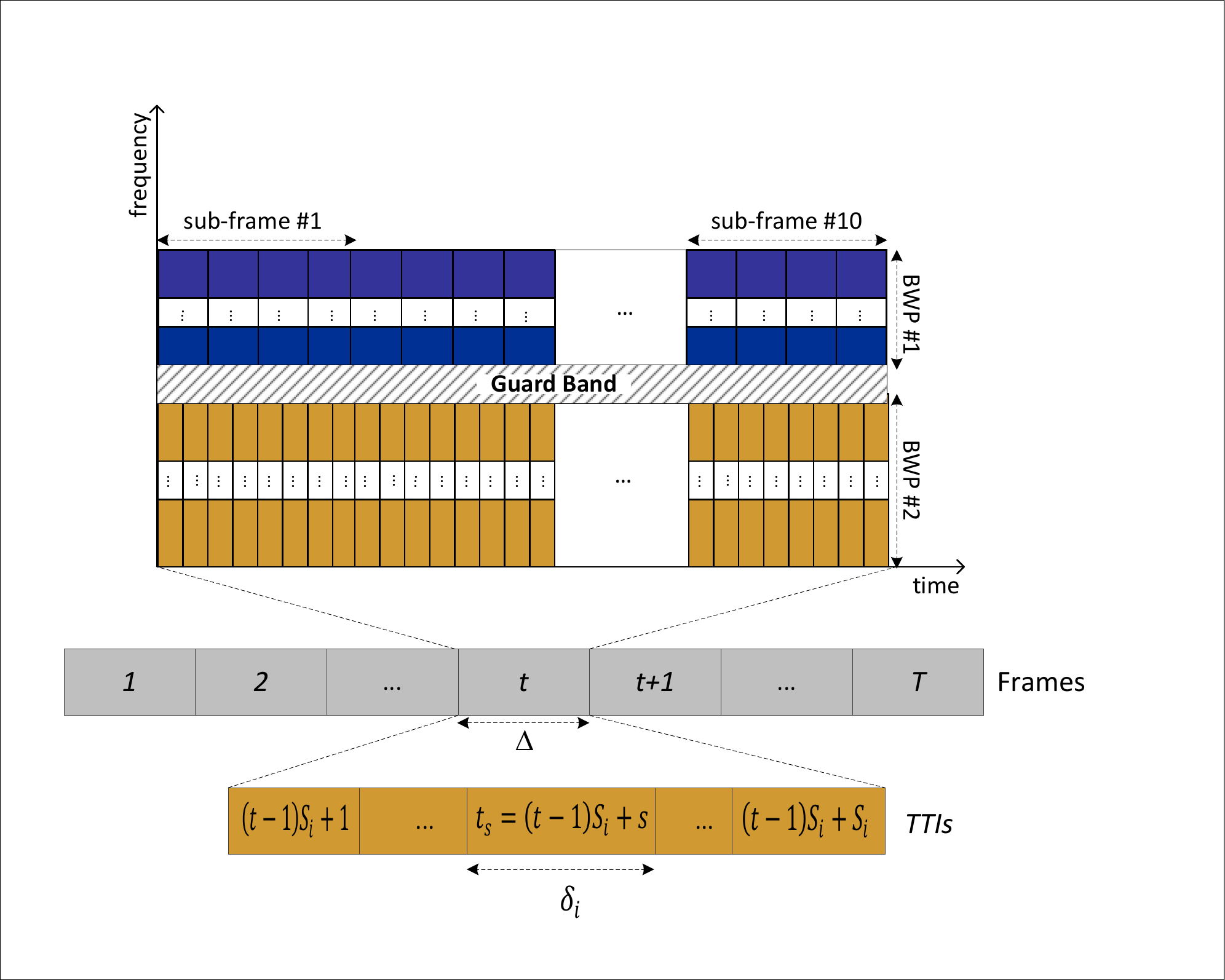}
    \caption{Time-frequency grid with different numerologies.}
    \label{fig3}
\end{figure}

The multiplexing of mixed numerologies in the frequency domain is considered in this work, where the carrier BW that is accessible for the downlink transmissions is divided into several bandwidth parts (BWPs). According to this, each user is able to alter its RF bandwidth based on its required data rate by switching between numerous BWPs. 
As illustrated in Fig. \ref{fig3}, the desirable BWP design to serve two types of services with different requirements is established based on the expected queue length of each service by introducing the BW-split variable $\alpha[t] \in [0,1]$. Whereas this method does not call for tight time synchronization techniques, using various numerologies in the adjacent sub-bands causes inter-numerology interference (INI). Hence, to reduce INI, a fixed guard band $B_{G}$ equal to one RB's BW (\textit{i.e.}, $180$ kHz) is configured between the two neighbour numerologies (\textit{i.e.}, sub-bands). The scheduled BWP assigned to uRLLC slice with numerology $i=2$ is denoted by $B_i[t]|_{i=2} = \alpha[t] B$, to unload the existing packets in the uRLLC slice's queues at frame $t$, where $B$ is the total carrier BW. In contrast, $B_i[t]|_{i=1} = (1-\alpha[t])B - B_{G}$ the scheduled BWP assigned to eMBB slice with numerology $i=1$. 

Assume the proposed system model works in a discrete time-frame indexed by $t \in [1, 2, \dots, T]$, which corresponds to one large-scale coherence time of $\Delta =10$ ms duration for each frame, as shown in Fig. \ref{fig3}. Depending on the selected numerology $i$ by each service, each frame in time-domain is divided into $S_i$ TTIs where duration of each TTI denoted by $t_s=(t-1)S_{i}+s$ with $s=\{1, \dots, S_{i}\}$ is $\delta_i $. Thus, based on the selected numerology $i$, each BWP is partitioned into $F_i$ number of sub-bands of frequency set $\mathscr{F}_i = \{1,
\dots, f_i, \dots ,F_i\}$ in the frequency-domain and $S_{i}$ number of TTIs in each frame, indexed by $t_s= \{(t-1)S_{i}+1, \dots , (t-1)S_{i}+s, \dots, (t-1)S_{i}+S_{i}\}$ in the time-domain. Such that, $F_i [t]= \lfloor{B_i[t]}/{\beta_i} \rfloor$ and $S_{i} = {\Delta}/{\delta_i}$. Therefore, a total $F_i[t]\times S_{i}$ number of RBs are accessible for the services using the $i$-th numerology at each frame $t$ via each RU. 

As depicted in Fig. \ref{fig2}, the $U$ independent data traffics with different demands at the CU layer are subsequently routed to VNFs in the DUs layer for parallel processing, referred to as data flows. We adopt the $M/M/1$ processing queue model on a first-come-first-serve basis to serve each user's packets. As it is clear from Fig. \ref{fig2}, the maximum number of paths for each user is $M$. According to the principle of the TS technique, the CU splits the data flow of the $u$-th user into several sub-flows, which are possibly transmitted via the maximum of $M$ paths and then aggregated at this user. Because of the non-overlapped DUs’ coverage, the resource optimization design at one DU is similar to that of other DUs. Thus, for ease of presentation, we drop the subscript index of DUs hereafter. To this end, we define $\boldsymbol{a}_u[t]\triangleq \big[a_{m,u}[t]\big]$ as the flow-split selection vector for the $u$-th data flow in time-frame $t$. In particular, if $a_{m,u}[t]=1$, the $m$-th RU is selected to transmit data of $u$-th data flow; otherwise, $a_{m,u}[t]=0$. In addition, let us denote by $\boldsymbol {\varphi} [t] \triangleq \{\boldsymbol{\varphi}_u [t],\ \forall u | \sum_{m}\varphi_{m,u}[t]=1, \varphi_{m,u}[t] \in [0,1]\}$ the global flow-split decision, in which $\boldsymbol{\varphi}_u [t] \triangleq \Big [\varphi_{m,u}[t] \Big]^T$ represents the flow-split portion vector of user $u$ while $\sum_{m}\varphi_{m,u}[t]=1$, where $\varphi_{m,u}[t] \in [0,1]$ indicates a portion of data flow routed to user $u$ via RU $m$ in time $t$ by selecting action $a_{m,u}[t]$.

\subsubsection{Achievable Throughput}
The channel vector between RU $m$ and the $u$-th user at the sub-band $f_i$ in TTI $t_s$ is denoted by $\boldsymbol{h}_{m,u,f_i}[t_s]\in \mathbb{C}^{K\times 1}$, which follows the Rician fading model with the Rician factor $\varrho_{m,u,f_i}[t]$. Within each frame, we assume that the channel remains temporally invariant, while it may be different across each short-time scale TTI. We model $\boldsymbol{h}_{m,u,f_i}[t_s]$ as:
\begin{align}\boldsymbol{h}_{m,u,f_i}[t_s] = &\ \sqrt{\zeta_{m,u,f_i}[t]}\big(\sqrt{\varrho_{m,u,f_i}[t]/(\varrho_{m,u,f_i}[t]+1)} \Bar{\boldsymbol{h}}_{m,u,f_i}[t]\nonumber\\
&+\sqrt{1/(\varrho_{m,u,f_i}[t]+1)}\tilde{\boldsymbol{h}}_{m,u,f_i}[t_s]\big)\end{align}
where $\zeta_{m,u,f_i}[t]$ is the large-scale fading; $\Bar{\boldsymbol{h}}_{m,u,f_i}[t]$ and $\tilde{\boldsymbol{h}}_{m,u,f_i}[t_s]$ are the line-of-sight (LoS) and non-LoS (NLoS) components, which follow a deterministic channel and Rayleigh fading model, respectively.  Given the orthogonality constraint, this work considers that each RB of a RU is assigned to only one single user during one TTI, such as $\pi_{m,u,f_i}^{\sfem}[t_s]\in\{0,1\}$ and $\pi_{m,u,f_i}^{\sfur}[t_s]\in\{0,1\}$ for eMBB and uRLLC traffics, respectively. Here, $\pi_{m,u,f_i}^{\sfem}[t_s]=1$ if the RB($t_s,f_i$) associated with sub-band $f_i$ in TTI $t_s$ of RU $m$ assigned to the $u$-th eMBB user, and $\pi_{m,u,f_i}^{\sfem}[t_s]=0$, otherwise; a similar definition is given for uRLLC users. Let define $\Pi [t_s]= \{\pi_{m,u,f_i}^{\sfem}[t_s], \pi_{m,u,f_i}^{\sfur}[t_s] \in \{0,1\}|\sum_{u}\big(\pi_{m,u,f_i}^{\sfem}[t_s]+\pi_{m,u,f_i}^{\sfur}[t_s]\big)\leq 1\}$ as the RB allocation constraint, which ensures the orthogonality constraint and QoS constraint for uRLLC service.

The instantaneous achievable rate in [bits/s] for a given set of channel realizations at the $u$-th eMBB user at TTI $t_s$ is given by:
\begin{equation}\label{eq1}
  r_{m,u}^{\sfem}(\boldsymbol{p}^{\sfem}[t_s])=\sum_{f_i=1}^{F_i}\beta_i \log_{2}\Bigl(1+\frac{p_{m,u,f_i}^{\sfem}[t_s] g_{m,u,f_i}[t_s]}{N_{0}}\Bigl),\ \forall u \in \mathscr{U}^{\sfem}
\end{equation}
where $\beta_i$, $N_0$ and $p_{m,u,f_i}^{\sfem}[t_s]$ are the bandwidth of each RB in numerology index $i$, power of the Additive White Gaussian Noise (AWGN), and  transmit power from RU $m$ to user $u$ for eMBB traffic at sub-band $f_i$ at the TTI $t_s$, respectively; $g_{m,u,f_i}[t_s] $ denotes the effective channel gain, given as $g_{m,u,f_i}[t_s] \triangleq \|\boldsymbol{h}_{m,u,f_i}[t_s]\|^2_2$.
 Let us define $\boldsymbol{p}^{\sfem}[t_s]\triangleq[p_{m,u,f_i}^{\sfem}[t_s]],\ \forall f_i, u, m$. The transmit power must satisfy $p_{m,u,f_i}^{\sfem}[t_s]\leq \pi_{m,u,f_i}^{\sfem}[t_s] P_m^{\max}$ with $P_m^{\max}$ being the power budget at  RU $m$, which guarantees that RU $m$ allocates power to user $u$ on RB$(t_s,f_i)$ only if $\pi_{m,u,f_i}^{\sfem}[t_s]=1$; otherwise $\pi_{m,u,f_i}^{\sfem}[t_s]=0$ and $p_{m,u,f_i}^{\sfem}[t_s]=0$.
As a result, the throughput of eMBB user $u \in \mathscr{U}^{\sfem}$ in TTI $t_s$ is given as $r_u^{\sfem}(\boldsymbol{p}^{\sfem}[t_s])=\sum_{m} r_{m,u}^{\sfem}(\boldsymbol{p}^{\sfem}[t_s])$. The minimum QoS requirement for eMBB users is guaranteed by the constraint $\sum_{t_s} r^{\sfem}_{u}{(\boldsymbol{p}^{\sfem}[t_s])}\geq R_{\mathsf{th}}$, where $R_{\mathsf{th}}$ is a given QoS threshold.

In contrast, owing to the finite block-length in uRLLC traffics, the instantaneous achievable rate of $u$-th uRLLC user from RU $m$ in TTI $t_s$ using the short block-length can be expressed as \cite{polyanskiy2010channel}:
\begin{align}\label{eq2}
    r_{m,u}^{\sfur}(\boldsymbol{p}^{\sfur}[t_s],\boldsymbol{\pi}^{\sfur}[t_s])= \sum_{f_i=1}^{F_i}&\beta_i \Big[\log_{2}\Big(1+\frac{ p_{m,u,f_i}^{\sfur}[t_s] g_{m,u,f_i}[t_s] }{N_{0}}\Big) \notag\\
    &-\frac{\pi_{m,u,f_i}^{\sfur}[t_s]\sqrt{V} Q^{-1}(P_{e})}{\sqrt{ \delta_i \beta_i}}\Big],\ \forall u \in \mathscr{U}^{\sfur}
\end{align}
where $V$, $P_{e}$ and $Q^{-1}$: $\{0,1\}\rightarrow \mathbb{R}$ denote the channel dispersion, error probability, and  inverse of the Gaussian Q-function, respectively. Let us define $\boldsymbol{p}^{\sfur}[t_s]\triangleq[p_{m,u,f_i}^{\sfur}[t_s]]$ and $\boldsymbol{\pi}^{\sfur}[t_s]\triangleq[\pi_{m,u,f_i}^{\sfur}[t_s]],\ \forall f_i, u, m$.
It is observed that $V=1-\frac{1}{(\Gamma[t_s])^2}\approx1$ when the received $\Gamma[t_s]=\frac{p_{m,u,f_i}^{\sfur}[t_s] g_{m,u,f_i}[t_s] }{N_{0}} \geq \Gamma_0$ with $\Gamma_0 \geq 5$ dB. This can be easily achieved in cellular networks, by arranging the uRLLC decoding vector into one possible null space of the reference subspace \cite{schiessl2015delay}. Hence, we consider the constraint $\frac{N_0 \Gamma_0}{g_{m,u,f_i}[t_s]} \pi_{m,u,f_i}^{\sfur}[t_s] \leq p_{m,u,f_i}^{\sfur}[t_s] \leq \pi_{m,u,f_i}^{\sfur}[t_s] P_m^{\max} $ to guarantee the approximation $V\approx 1$ as well as the big-$M$ formulation theory to avoid non-convexity of (\ref{eq2}). Similar to the eMBB service, the throughput of uRLLC user $u \in \mathscr{U}^{\sfur}$ in TTI $t_s$ is given as $r_u^{\sfur}(\boldsymbol{p}^{\sfur}[t_s],\boldsymbol{\pi}^{\sfur}[t_s])=\sum_{m} r_{m,u}^{\sfur}(\boldsymbol{p}^{\sfur}[t_s],\boldsymbol{\pi}^{\sfur}[t_s])$. We have the following power constraint as:
\begin{equation}\label{eq3}
\begin{split}
\mathscr{P}[t_s] = &\Big\{ 0 \leq p_{m,u,f_i}^{\sfem}[t_s]\leq \pi_{m,u,f_i}^{\sfem}[t_s] P_m^{\max},\\
&\frac{N_0 \Gamma_0}{g_{m,u,f_i}[t_s]} \pi_{m,u,f_i}^{\sfur}[t_s] \leq p_{m,u,f_i}^{\sfur}[t_s] \leq \pi_{m,u,f_i}^{\sfur}[t_s] P_m^{\max}|\\
&\sum_{i}\sum_{f_i,u}^{}(p_{m,u,f_i}^{\sfem}[t_s]+p_{m,u,f_i}^{\sfur}[t_s])\leq P_{m}^{\max}\Big\}.
\end{split}
\end{equation}

We denote $\lambda_u[t]$ in [packets/s] as the unknown traffic demand of user $u$ in time-frame $t$ with the length of $Z^{\mathsf{x}}$ bytes with $\mathsf{x} \in \{\sfur,\sfem\}$, which is i.i.d. over time and  upper bounded by a finite constant $\lambda^{\max}$, such as $\lambda_u[t] \leq \lambda^{\max} \leq \infty$. We consider that the retained independent queue at each RU for the $u$-th user, which is denoted by $\{\varphi_{m,u}[t] \lambda_u[t] Z^{\mathsf{x}}\}$ as the arrival processes of sub-flows, is controlled by a congestion scheduler.  Thus, the queue-length of data flow $u$ at RU $m$ in TTI $(t_{s+1})$ is $q_{m,u}[t_{s+1}] = \max\{\big [q_{m,u}[t_s] + \varphi_{m,u}[t] \lambda_u[t] Z^{\mathsf{x}} \Delta - r_{m,u}^{\mathsf{x}} [t_s]\delta_i\big ],0\}$. In order to avoid the packet loss due to buffer overflow in each RU, the constraint $\sum_{u} q_{m,u}[t_s] \leq Q^{\max}, \forall m$ is imposed to ensure that the available packets in the buffer of RU shouldn't exceed the maximum queue-length of $Q^{\max}$ for each RU. Let $\boldsymbol{q}[t_s] \triangleq \big[q_{m,u}[t_s] \big]^T,\ \forall m, u$. \\

\subsubsection{The E2E Traffic Latency for uRLLC}
Denote by $f_{cu}$ and $f_{du}$ the computation capacities of CU and DU [cycles/sec], respectively. Considering the identical packet size, the required computation resource to process one packet of size $Z$ is $C$ (number of cycles). As result, $\mu_{cu}=f_{cu}/C$ and $\mu_{du}=f_{du}/C$ are the task rates [1/sec] at CU and DU, respectively. As a result,  $1/\mu_{cu}$ and $1/\mu_{du}$ represent the mean service time of  CU and  DU layers, respectively. The processing latency of all data flows at the CU layer ($\tau^{\pro}_{cu}$) and DU layer ($\tau^{\pro}_{du}$) is computed as:
\begin{align}\label{eq4}
    \nonumber \tau^{\pro}_{cu}[t]=\frac{\Lambda[t]}{\mu_{cu}},\ \text{and}\
    \tau^{\pro}_{du}[t]=\frac{\Lambda[t]}{\mu_{du}},\ \forall n \in \mathscr{N}
\end{align}
where $\Lambda[t] = \sum_u \lambda_u[t]$.
Next, the arrival packets $\lambda_{u}[t]$ for the $u$-th user is transported to the DU layer via the midhaul (MH) link with the maximum capacity $C^{\mathsf{MH}}$ [bits/sec] between CU and DU. By Burke's theorem, the mean arrival data rate of the second layer, which is processed in the first layer, is still the same rate \cite{burke1956output}. Hence, the data transmission latency of the traffic flow for user $u$ under the MH limited capacity is:
\begin{equation}\label{eq6}
    \tau^{\tx}_{cu,du} [t]=\frac{\Lambda[t]Z}{C^{\mathsf{MH}}}.
\end{equation}

As mentioned previously, the maximum number of paths from DU $n$ to each user is $M_n$. Since the packets for user $u$ can be transmitted by multiple RUs, the effective response time $\tau_{du,ru}^{\tx}$ to transport all packets the DUs layer should be computed by the worst average response time among its connected FH links with maximum capacity $C^{\mathsf{FH}}_{m}$ [bits/sec], \textit{i.e.},
\begin{align}\label{eq7}
    \tau_{du,ru}^{\tx} [t]= \max_{m}\Big\{\frac{\sum_{u\in\mathscr{U}^{\sfur}}\varphi_{m,u}[t]\lambda_u[t] Z^\sfur}{C^{\mathsf{FH}}_{m}}\Big \},\ \forall m\in \mathscr{M}_n.
\end{align}
The transmission latency from RU $m$ to user $u$ is then calculated as:
\begin{equation}\label{eq8}
\tau_{ru,u}^{\tx} [t_s]= \max_{m}\Big\{\frac{\varphi_{m,u}[t]\lambda_u[t] Z^\sfur}{r_{m,u}^{\sfur}[t_s]}\Big\},\    \forall u\in \mathscr{U}^{\sfur}.
\end{equation}
Simply put, the e2e latency of each uRLLC user $u\in \mathscr{U}^{\sfur}$ per each TTI is computed as:
\begin{align}\label{eq9} 
\tau_{u}^{\sfur} [t]=\tau^{\pro}_{cu}[t]+\tau^{\tx}_{cu,du}[t]+\tau^{\pro}_{du}[t]+\tau_{du,ru}^{\tx}[t]
+\sum_{t_s}\big(\tau_{ru,u}^{\tx}[t_s]+ \tau^{\pro}_{ru}[t_s] \big),\ \forall u\in \mathscr{U}^{\sfur}
\end{align}
where $\tau^{\pro}_{ru}$ is the process latency at RU $m$, which is bounded by three OFDM symbols duration that is typically very small. To ensure a minimum latency requirement for uRLLC user $u$, the e2e latency is bound by a predetermined threshold $D_u^{\sfur}$, \textit{i.e.}, $\tau_u^{\sfur} [t]\leq D_u^{\sfur}$.

\section{Problem Formulation and Overall Intelligent \\ Traffic Steering Algorithm} \label{section3}

\subsection{Problem Formulation}
\noindent \textbf{Utility function:} The ultimate goal is to optimize the joint intelligent traffic prediction, flow-split distribution, dynamic user association and radio resource management in the presence of unknown dynamic traffic demand to serve eMBB and uRLLC users, subject to various resources constraints and diverse QoS requirements. Due to the conflict of objective functions in both services (\textit{i.e.} eMBB and uRLLC), the utility function should capture the eMBB throughput and worst-user e2e uRLLC latency separately such as: $\mathcal{R}^{\sfem}=\sum_{u \in \mathscr{U}^{\sfem}} r_u^{\sfem}(\boldsymbol{p}^{\sfem}[t_s])$ and $\max_{u\in\mathscr{U}^{\sfur}}\{\tau_{u}^{\sfur}\}$ on two independent optimization problems. Based on the above definitions and discussions, the JIFDR problem is mathematically formulated as the two independent optimization problems with common constraints as follows: 
\begin{subequations}\label{op:1}
\begin{IEEEeqnarray}{cl}
 \text{P1}:\ \max_{\boldsymbol{\lambda},\boldsymbol{\varphi},\boldsymbol{\pi},\boldsymbol{p}, \alpha}&\quad  \mathcal{R}^{\sfem}(\boldsymbol{p}^{\sfem}[t_s])  \label{op:1a}\\ 
 \text{s.t.}  \quad &\quad \boldsymbol{\pi}[t_s]\in \Pi[t_s], \ \forall t_s\label{op:1b}\\
  &  \quad  \boldsymbol{p}[t_s] \in \mathscr{P}[t_s], \ \forall t_s \label{op:1c}\\
  &  \quad \boldsymbol{\varphi}_u [t] \in \boldsymbol{\varphi}[t],\   \forall t, u\in \mathscr{U}\label{op:1d}\\
  &   \quad   \sum_{t_s}r^{\sfem}_{u}(\boldsymbol{p}^{\sfem}[t_s])\geq R_{\mathsf{th}}, \ \forall u\in \mathscr{U}^{\sfem}\label{op:1e}\\
   &   \quad \sum_u^{}\big[r_{m,u}^{\sfem}(\boldsymbol{p}^{\sfem}[t_s])+r_{m,u}^{\sfur}(\boldsymbol{p}^{\sfur}[t_s],\boldsymbol{\pi}^{\sfur}[t_s])\big]\leq C_{m}^{\mathsf{FH}},\ \forall m\in \mathscr{M}_n\label{op:1f}\\
  &  \quad \sum_{t_s} r_{m,u}^{\sfur}(\boldsymbol{p}^{\sfur}[t_s],\boldsymbol{\pi}^{\sfur}[t_s])\geq \frac{\varphi_{m,u}[t]\lambda_{u}[t] Z^{\sfur}} {\Delta}, \ \forall m\in \mathscr{M}_n, u\in \mathscr{U}^{\sfur}\label{op:1g}\\
  & \quad \tau^{\sfur}_{u} (\boldsymbol{\lambda}[t],\boldsymbol{\varphi}[t],\boldsymbol{\pi}[t_s],\boldsymbol{p}[t_s])\leq D^{\sfur}_u, \  \forall u\in \mathscr{U}^{\sfur}\label{op:1h}\\
 & \quad \sum_{u} q_{m,u}[t_s] \leq Q^{\max}, \ \forall t_s, m\in \mathscr{M}_n \label{op:1i}\\
  & \quad \sum_{f_i=1}^{F_i} \beta_i \leq B_i[t],\ i \in\{1,2\}\label{op:1j}\\
  & \quad 0 \leq \alpha[t] \leq 1 \label{op:1k}
\end{IEEEeqnarray}
\end{subequations}
and
\begin{subequations}\label{op:2}
\begin{IEEEeqnarray}{cl}
 \text{P2}:\ \min_{\boldsymbol{\lambda},\boldsymbol{\varphi},\boldsymbol{\pi},\boldsymbol{p}, \alpha}&\quad \max_{u\in \mathscr{U}^{\sfur}}\{\tau_{u}^{\sfur}\}  \\ 
 \text{s.t.} \quad & \eqref{op:1b}-\eqref{op:1k}
\end{IEEEeqnarray}
\end{subequations}
where $\boldsymbol{\varphi}[t],\boldsymbol{\pi}[t_s]$ and $\boldsymbol{p}[t_s]$ are the vectors encompassing the flow-split portions, sub-band assignments and power allocation variables at frame $t$ and TTI $t_s$, respectively. Recall that, for each BWP with the given numerology, $B_i[t]|_{i=2} =\alpha[t] B$ and $B_i[t]|_{i=1} = (1-\alpha[t])B- B_{G}$. Constraint \eqref{op:1f} expresses the limited capacity of FH link between DU $n$ and RU $m$. Constraint \eqref{op:1g} ensures that each RB assigned to the $u$-th uRLLC user should transmit a complete data packet with the size $Z^{\sfur}$.

\textbf{Challenges of solving JIFDR problem}:  
The main challenges in solving problems (P1) and (P2) lie in the non-convexity of $\tau^{\sfur}_u$ and constraints \eqref{op:1f}, \eqref{op:1g} and \eqref{op:1i} with respect to flow-split portions and transmit power variables. Furthermore, the binary nature of sub-band allocation variables in constraint \eqref{op:1b} makes these problems more difficult to solve directly, which is generally MINCP. Once may employ the MINCP solvers (\textit{e.g.} Gurobi) to directly solve binary $\boldsymbol{\pi}$. However, we argue that the exponential computation complexity of such a MINCP formulation limits its practical feasibility, especially when the number of variables exceeds few thousand in large-scale scenarios. Besides, the traffic demand $\lambda[t]$ for the next time (frame) is unknown in practice. Such that the BW-split $\alpha[t]$ and flow-split vectors $\varphi[t]$ for frame $t$ will be decided based on the previous states updated by the RAN layer and knowledge of the previous traffic demands $\{\lambda[t-1]\}_{\forall t}$. In order to attain high QoE for all users in each TTI, an efficient and adaptable solution to the long-term subproblem of (\ref{op:1}) and (\ref{op:2}) is required. 

\subsection{Sub-Optimization Problems}
It is clear, both problems (\ref{op:1}) and (\ref{op:2}) must be solved on separate time scales, \textit{i.e.} on the long-term scale $t$ and the short-term scale $t_s$. To reduce the computational complexity and information sharing as well as to provide a stable queuing system, the traffic demand vector $\boldsymbol{\lambda} [t]$, the flow-split decision vector $\boldsymbol{\varphi}[t]$ and BW-splitting variable 
$\alpha [t]$ are only solved and updated once per time-frame $t$. In contrast, the power allocation vector $\boldsymbol{p}[t_s]$ and the RB allocation vector $\boldsymbol{\pi}[t_s]$ are optimized in every TTI $t_s$, adapting to dynamic environments. Since both P1 and P2 have the same condition, a similar procedure is applied to solve both. From now on, we take consider only the P1 in detail.  
\subsubsection{Long-term Subproblem (L-SP)} 
The joint optimization subproblem of the traffic demand, flow-split distribution and dynamic RAN slicing at time-scale $t$ is re-expressed as:
\begin{subequations}\label{sp:1}\begin{IEEEeqnarray}{cl}
\text{L-SP}: \max_{\boldsymbol{\lambda},\boldsymbol{\varphi}, \alpha}&\quad  \mathcal{R}^{\sfem}(\boldsymbol{p}^{\sfem}[t_s]) \label{sp:1a} \\ 
 \text{s.t.} \quad & \quad \boldsymbol{\varphi}_u [t] \in \boldsymbol{\varphi}[t],\   \forall t, u\label{sp:1b}\\
  &  \quad \sum_{t_s} r_{m,u}^{\sfur}(\boldsymbol{p}^{\sfur}[t_s],\boldsymbol{\pi}^{\sfur}[t_s])\geq \frac{\varphi_{m,u}[t]\lambda_{u}[t] Z^{\sfur}} {\Delta},  \ \forall  m, u\label{sp:1c}\\
  & \quad \tau^{\sfur}_{u} (\boldsymbol{\lambda}[t],\boldsymbol{\varphi}[t],\boldsymbol{\pi}[t_s],\boldsymbol{p}[t_s])\leq D^{\sfur}_u, \  \forall u\label{sp:1d}\\
  & \quad \sum_{f_i=1}^{F_i} \beta_i \leq B_i[t],\ i \in\{1,2\}\label{sp:1e}\\
  & \quad 0 \leq \alpha[t] \leq 1.\label{sp:1f}
\end{IEEEeqnarray}
\end{subequations}
Although the L-SP (\ref{sp:1}) is non-convex due to non-convexity of constraints \eqref{sp:1c} and \eqref{sp:1d}, it cannot be solved directly by standard optimization techniques because $\boldsymbol{\lambda}[t]$ is completely unknown at the beginning of each frame. At the next section, three successive methods are proposed for solving this problem, that predict traffic demand, dynamic BW-split distribution and dynamic flow-split variables as $\boldsymbol{\lambda}^*[t]$, $\boldsymbol{\alpha}^*[t]$ and $\boldsymbol{\varphi}^*[t]$ at the beginning of each frame $t$, respectively.

\subsubsection{Short-term Subproblem (S-SP)} Given $\boldsymbol{\lambda}^*[t]$, $\boldsymbol{\alpha}^*[t]$, and $\boldsymbol{\varphi}^*[t]$ forwarded from the non-RT RIC through the A1 interface, the resource allocation problem at time slot $t_s$ in the near-RT RIC is expressed as:
\begin{subequations}\label{sp:2}\begin{align}
\text{S-SP}: \max_{\boldsymbol{\pi},\boldsymbol{p}}&\quad  \mathcal{R}^{\sfem}(\boldsymbol{p}^{\sfem}[t_s])  \\ 
 \text{s.t.} \quad & \quad \boldsymbol{\pi}[t_s]\in \Pi[t_s], \ \forall t_s \label{sp:2b}\\
  & \quad  \boldsymbol{p}[t_s] \in \mathscr{P}[t_s], \ \forall t_s  \label{sp:2c}\\
  &  \quad   \sum_{t_s}r^{\sfem}_{u}(\boldsymbol{p}^{\sfem}[t_s])\geq R_{\mathsf{th}}, \  \forall u \label{sp:2d}\\
   &\quad \sum_u^{}\big[r_{m,u}^{\sfem}(\boldsymbol{p}^{\sfem}[t_s])+r_{m,u}^{\sfur}(\boldsymbol{p}^{\sfur}[t_s],\boldsymbol{\pi}^{\sfur}[t_s])\big]\leq C_{m}^{\mathsf{FH}},\ \forall m\label{sp:2e}\\
  &  \quad \sum_{t_s} r_{m,u}^{\sfur}(\boldsymbol{p}^{\sfur}[t_s],\boldsymbol{\pi}^{\sfur}[t_s])\geq \psi,  \ \forall m, u \label{sp:2f}\\
  &\quad \tau^{\sfur}_{u} (\boldsymbol{\pi}[t_s],\boldsymbol{p}[t_s])\leq D^{\sfur}_u, \  \forall u\\
 & \quad \sum_{u} q_{m,u}[t_s] \leq Q^{\max}, \ \forall t_s, m\in \mathscr{M}_n  \label{sp:2h}
 \end{align}
 \end{subequations}
where $\psi = \frac{\varphi^*_{m,u}[t] \lambda_{u}^*[t] Z^{\sfur}} {\Delta}$. The S-SP \eqref{sp:2} involves both binary ($\boldsymbol{\pi}$) and continuous ($\boldsymbol{p}$) optimization variables with nonlinear objective function and non-convex constraint \eqref{sp:2e} at time slot $t_s$, which is still remained an MINCP problem. Since MINCP problems incorporate the optimizing challenges under integer variables with managing nonlinear functions, such problems comprise an immense class of difficult optimization problems.

\subsection{Overall Intelligent Traffic Steering Deployment Architecture and Algorithm}
\begin{figure}
\centering
    \includegraphics[width=0.75\textwidth,trim=2 2 2 2,clip=true]{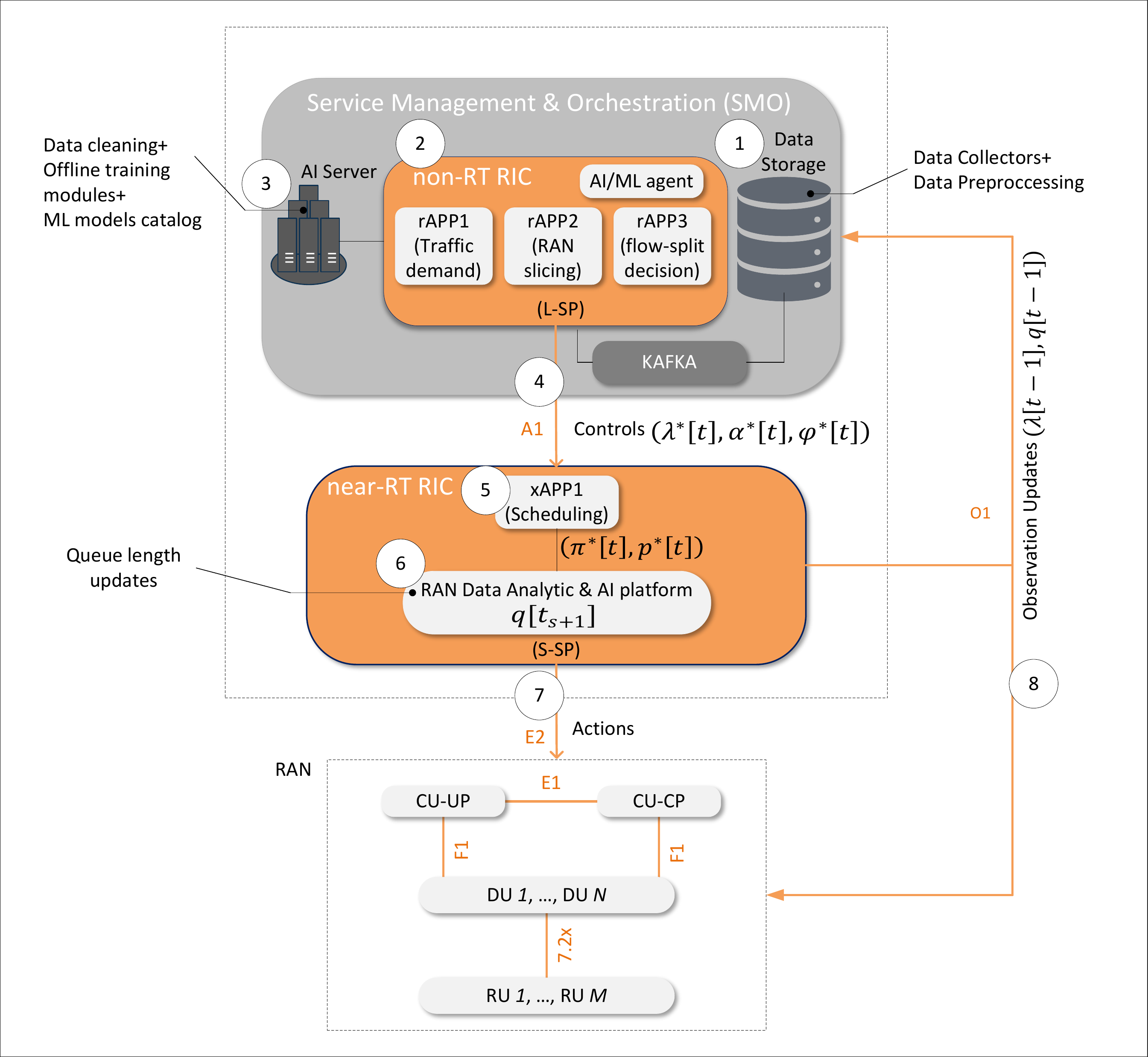}
    \caption{High-level structure of deploying the proposed intelligent traffic prediction and JIFDR management scheme within the ORAN architecture.}
    \label{fig4}
\end{figure}

In Fig. \ref{fig4}, we show the high-level organization of deployment scenarios and the end-to-end flow of the proposed algorithm within the ORAN architecture. This is inspired by the second set of deployment scenarios listed in the technical report \cite{alliance2019ran} by the ORAN Alliance.
\begin{enumerate}[label=\protect\circled{\arabic*}]
    \item The collected data, including performances/observations and resource updates from RAN components and near-RT RIC, are collected into a data collector located at the SMO. This process is done via the O1 interface. Based on these collected data in SMO, three rAPPs for solving  L-SP are carried out at non-RT RIC. For $t=1$, we assume a random traffic demand with Poisson process and equal flow-split decision for all paths.
    
    \item Utilizing a data bus like Kafka, the collected data at the SMO is routed to non-RT RIC  in the SMO.
    
    \item The non-RT RIC queries the relevant ML/AI model, which is hosted in the AI server within the SMO. Once the model has been well trained on the AI server, non-RT RIC is notified of the inference.
    
    \item The scheduling xAPP in near-RT RIC is then loaded with inference results and policies via the A1 interface. Applications, that are designed specifically for radio functions or xAPPs, enable  RAN components to be programmed.
    
    \item Given ${\boldsymbol{\lambda}}^*[t]$, $\boldsymbol{\alpha}^* [t]$ and $\boldsymbol{\varphi}^*[t]$, xAPP1 deployed in near-RT RIC controls congestion through MC technique and optimizes RAN resources and functions in each time-slot $t_s$ by solving S-SP to obtain optimal solutions of RB allocation $\boldsymbol{\pi}^*[t_s]$ and power allocation $\boldsymbol{p}^*[t_s]$.
    
    \item Subsequently, the RAN Data Analytic component in near-RT RIC updates queue lengths.
    
    \item Through the E2 interface, the relevant solution is transferred to CU or DU layers.
    
    \item After $S_i$ TTI (\textit{i.e.} one frame), the performance and observations (\textit{e.g.} $\boldsymbol{q}[t-1]$, $\boldsymbol{\lambda}[t-1]$) are updated to SMO through the O1 interface to re-estimate the traffic demand ${\boldsymbol{\lambda}}^*[t+1]$ and flow-split decision $\boldsymbol{\varphi}^*[t+1]$.
\end{enumerate}

The overall intelligent TS algorithm to solve the JIFDR problem \eqref{op:1} is summarized in Algorithm \ref{alg1}, where the solutions for subproblems will be detailed in Section \ref{section4}. It is straightforward to develop a similar procedure to solve problem \eqref{op:2}.

\begin{algorithm}[t]
\begin{algorithmic}[1]
\fontsize{11}{911}
\protect\caption{Proposed Intelligent Traffic Steering Algorithm to Solve JIFDR Problem (\ref{op:1})}
\label{alg1}
\global\long\def\algorithmicrequire{\textbf{Initialization:}}
\REQUIRE  Set $t=1$, $t_s=1$, $\varphi_u[1]= \frac{1}{M}[1, \dots ,1]$ and $\alpha[1]=\frac{1}{2}$; all initial queues are set to be empty $q_{m,u}[1]=0$ and $\boldsymbol{q}[1]=0$. 
\FOR {$t=1,2, \dots,T$} 
\STATE \texttt{\textbf{Traffic demand prediction:}} Given ($\boldsymbol{\lambda}[t-1]$, $\boldsymbol{q}[t-1]$), non-RT RIC splits the available of all RUs' BW and traffic flows of all users by  (\ref{eq10}) and (\ref{eq11}) based on the predicted traffic demand (or arrival data rate) ${\boldsymbol{\lambda}}^*[t]$ by solving the L-SP \eqref{sp:1}
    \FOR {$t_s=1,2, \dots,S_i$ with $s \in \{1,2, \dots, S_i\}$}
    \STATE \texttt{\textbf{Optimizing scheduling:}} Given the queue-length vector $\boldsymbol{q}[t_s]$, and all long-term variables such as (${\boldsymbol{\lambda}}^*[t], \boldsymbol{\alpha}^*[t]$, and $\boldsymbol{\varphi}^*[t]$), solve the problem (\ref{op:3}) by Algorithm \ref{alg2} to obtain the RB assignment ($\boldsymbol{\pi}^*$) and power allocation ($\boldsymbol{p}^*$)
    \STATE \texttt{\textbf{Updating queue-lengths:}} Queue-lengths are updated as
    \[q_{m,u}[t_{s+1}] = \max\{\big [q_{m,u}[t_s] + \varphi_{m,u}[t] \lambda_u[t] Z^{\mathsf{x}} \delta_i - r_{m,u}^{\mathsf{x}} [t_s]\delta_i\big ],0\}\]
    where $\mathsf{x} \in \{\sfur, \sfem\}$.
    \STATE Set $s=s+1$
    \ENDFOR
\STATE Update $\{\boldsymbol{q}[t],\boldsymbol{\lambda}[t]\} = \{q_{m,u}[t], \lambda_u[t]\},\ \forall u \in \mathscr{U},\ m \in \mathscr{M}_n$
\STATE Set $t=t+1$
\ENDFOR 
\end{algorithmic}
\end{algorithm}

\section{Proposed Frameworks for Solving Subproblems} \label{section4}
We are now in a position to solve the L-SP and S-SP on different time scales. The optimal solutions for all optimization variables ($\alpha$ $\boldsymbol{\varphi}$, $\boldsymbol{\pi}$ and $\boldsymbol{p}$) strongly depend on the traffic demand vector $\boldsymbol{\lambda}$, which often require prior knowledge of the actual traffic of all services at non-RT RIC. Moreover, due to the dynamic environment and data collected from the RAN layer being only updated to non-RT RIC on a long-term scale, the assumption of complete information is unrealistic. In this paper, we aim to leverage observable historical system knowledge gathered over previous time-slots via the O1 interface to build a smoother optimal response to maximize the long-term utility.

\subsection{LSTM for Solving L-SP}

As mentioned previously, the L-SP cannot be solved directly by standard optimization techniques since $\boldsymbol{\lambda}[t]$ and  $\boldsymbol{q}[t_s]$ are often unknown at the beginning of each frame. Besides, the main challenge in optimizing traffic steering is to predict traffic precisely before the beginning of the next frame. An optimal policy cannot be implemented with an imprecise prediction of future traffic. In this section, utilizing a deep learning approach, we develop a data-driven real-time traffic demand prediction method. We suppose that the queue length of data flow $u$ in the next frame will depend on the traffic demand of data flow $u$ in the current and previous ones. Basically, RNN models utilize the current input as well as the output of one layer as the input for the subsequent layer. In such models, each layer is fed by the very first layer's input. This allows the RNN model to learn from the current and former time steps and then provides more precise predictions for traffic flows. These standard RNN models suffer from short-term memory owing to the vanishing and exploding gradient problems, which appear with longer data sequences. Due to these difficulties, the gradient either entirely disappears or explodes to a very high value, which makes them difficult to learn some long-period dependencies. To address the long-term dependency issue, the LSTM model has seen extensive use in the field of traffic prediction due to its capabilities in dealing with the long time-series flow data. As a result, we utilize the LSTM RNN to learn and predict the traffic pattern of all users in the considered ORAN architecture. 
\begin{figure}[t]
\centering
    \includegraphics[width=1 \textwidth,trim=2 2 2 2, clip=true]{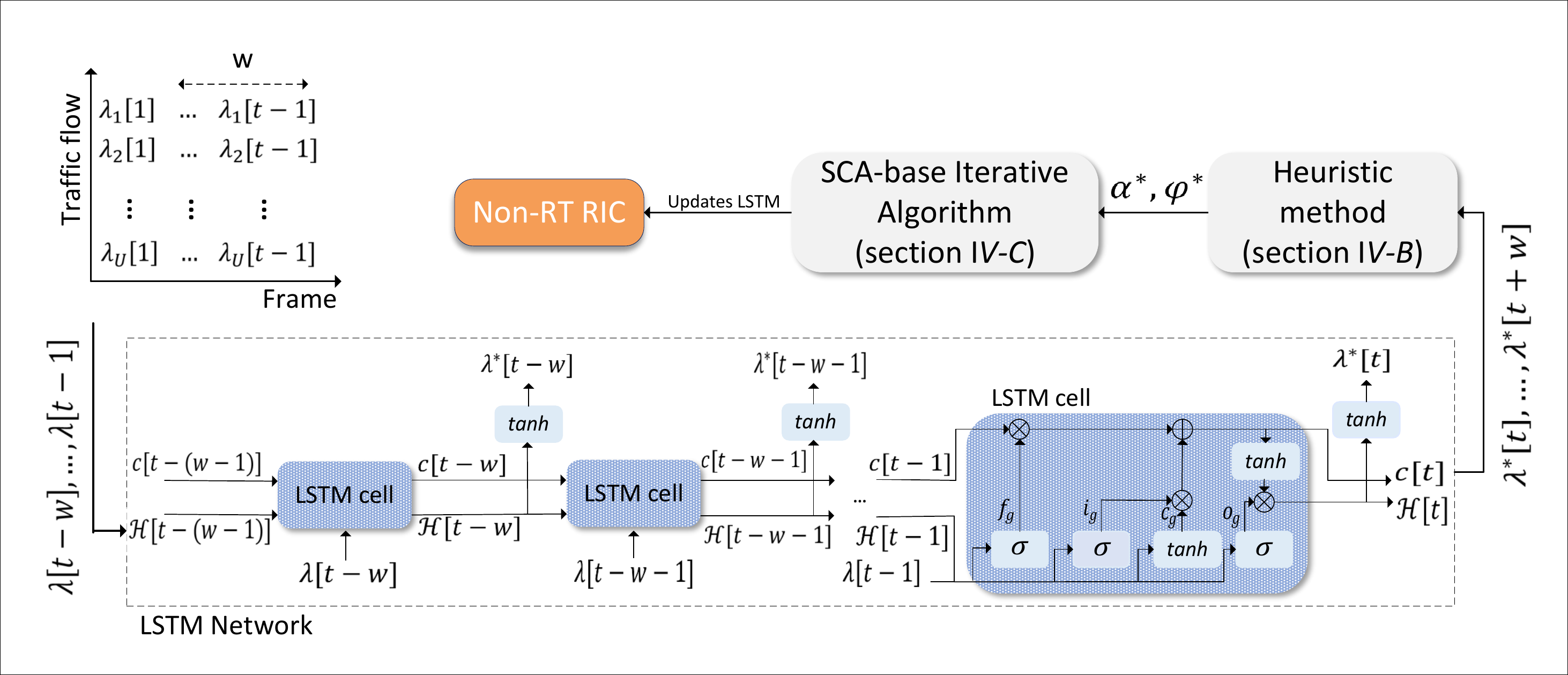}
    \caption{Implementing the proposed JIFDR management scheme at time-frame $t$.}
    \label{fig5}
\end{figure}

The fact that LSTM includes a memory cell to keep observable data, allowing them to handle long-term time series. As shown in Fig. \ref{fig5}, the structure of standard LSTM cells learns through four main gates, \textit{namely} input ($i_g$), forget ($f_g$), cell state-update ($c_g$) and output ($o_g$), that allows the input data to pass from the previous cells in the learning procedure. The output calculated by the input gate ($i_g$) and the cell state update ($c_g$) modify the current cell's state ($c[t]$), while the forget gate enables the current cell to discard or preserve the previous state value. To determine this, we take into account the output of the previous hidden state ($\mathcal{H}[t-1]$) and the actual input data ($\boldsymbol{\lambda}[t-1]$). The new cell state's value is based on the actual input and previous output of the cell. In contrast to other gates that employ the Sigmoid function, the cell state update benefits the hyperbolic tangent as an activation function that yields values between ${-1}$ and $1$. Eventually, the input, forget and cell state update gates are combined to create the current cell state. The current cell's output is determined as a function of the previous timestep's output ($\mathcal{H}[t-1]$), the actual input data ($\boldsymbol{\lambda}[t-1]$), and the cell state ($c[t-1]$) through the output gate. Lastly, after crossing through an activation function, the prediction value is calculated. Each LSTM layer comprises a chain of LSTM cells, in which the computed operation of each cell is transmitted to the next cell as an input. As illustrated in Fig. \ref{fig5}, the temporal pattern of the mentioned parameter are learned through the current and a window of previous traffic demands value with the length $W$ \{$\boldsymbol{\lambda[t-W]}, \boldsymbol{\lambda[t-W+1]}, \dots, \boldsymbol{\lambda[t-1]}$\} to predict future values.

The LSTM model is trained at non-RT RIC in the ORAN architecture, using long-term data gathered from RAN via O1. The near-RT RIC of the ORAN is then given access via the A1 interface to the trained model for inference. Upon the inference outcome, the intelligent TS is applied through the MC technique to enhance the associated key performance indicators (KPIs). Traffic demand prediction and the corresponding intelligent TS schemes are continually implemented till the desired KPI values, or the required QoS of traffic are met. In the following, the network parameter of data arrival rate $\boldsymbol{\lambda}$ is continuously monitored across all cells of RUs. Upon predicting the data arrival rate per frame, the flow-split distribution, dynamic RAN slicing and radio resource management with the MC technique can be applied to steer data flows. The weights of the RNN model are eventually updated depending on the actual parameter's value to reflect changes and enhance the performance till the goal KPI criteria are met if the prediction outcome is incorrect.

\subsection{Heuristic Methods for Predicting $\boldsymbol{\alpha}[t]$ and $\boldsymbol{\varphi}[t]$}

Upon the inference outcome of the LSTM model, the predicted traffic demands at the next frame $\boldsymbol{\lambda}^*[t]$ are transmitted immediately to two other embedded rAPPs in non-RT RIC for optimizing the dynamic bandwidth separation, $\boldsymbol{\alpha}[t]$ and flow-split decisions, $\boldsymbol{\varphi}[t]$. For efficient deployment, these parameters are designed in a longer time scale, \textit{i.e.}, on the frame basis compared to the time slot basis of power allocation and resource block assignment. Therefore, at the beginning of each frame, $\boldsymbol{\alpha}[t]$ and $\boldsymbol{\varphi}[t]$ should be determined upon getting the predicted traffic demands. Having optimum values of the bandwidth separation and flow split is very difficult if not possible because of the unknown CSI of future time slots in the current frame. Therefore, we propose an efficient heuristic algorithm to determine  $\boldsymbol{\alpha}[t]$ and $\boldsymbol{\varphi}[t]$ based on $\boldsymbol{\lambda}^*[t]$. An intuitive way is to allocate the bandwidth to each service proportionally to the corresponding traffic demands. However, since the amount of uRLLC traffics is much smaller than the amount of eMBB traffics, this method is not efficient in meeting the stringent latency requirement of uRLLC applications. To tackle this, we incorporate the maximum tolerable delays of both services and the total traffic demands. Thus, the bandwidth separation between eMBB and URLLC services is computed as follows: 
\begin{equation}\label{eq10}
\alpha^{*}[t] = \frac{\sum_{\mathscr{U}^{\sfur}}{\lambda}_u^*[t]}{\sum_{\mathscr{U}^{\sfem}}{\lambda}_u^*[t]}\times \frac{\tau^{\sfem}_{th}}{\tau^{\sfur}_{th}}
\end{equation}
where $\tau^{\sfur}_{th}$ and $\tau^{\sfem}_{th}$ represent the maximum allowed latency for uRLLC and eMBB services, respectively.
To plan the flow splitting factor $\varphi_u[t]$, we consider each DU's capacity in delivering user traffic demands $u$. Because we do not know the data rate for the user in the next frame, we take the moving average of the rate in the most recent time slots. For a generic user $u$ (can be uRLLC or eMBB user), let us define $Z_{m,u}[t]= \frac{1}{W} \sum_{l=t-W+1}^{t} r_{m,u}[l]$, where $r_{m,u}[l]$ is the achievable rate of user $u$ served RU $m$ at time slot $l$, and $W$ is the window size. The flow split for user $u$ to RU $m$ is computed as follows:
\begin{equation}\label{eq11}
\varphi^{*}_{m,u}[t] = \frac{Z_{m,u}[t]}{\sum_{m\in \mathcal{M}_n} Z_{m,u}[t]},\ \forall m,u. 
\end{equation}

\subsection{SCA-based Iterative Algorithm for solving S-SP}
To solve  problem (\ref{sp:2}) as a MINCP, we first relax binary variables to continuous ones (\textit{i.e.} the box constraints between 0 and 1) and transform constraint (\ref{sp:2e}) into a more traceable form which the SCA-based iterative algorithm can efficiently solve. 
\begin{algorithm}[!h]
\begin{algorithmic}[1]
\fontsize{12}{12}
\protect\caption{The Proposed SCA-based Iterative Algorithm to Solve S-SP (\ref{op:3})}
\label{alg2}
\global\long\def\algorithmicrequire{\textbf{Initialization:}}
\REQUIRE  Set $j:=0$ and generate initial feasible points for $(\boldsymbol{\pi}^{(0)}[t_s],\boldsymbol{p}^{(0)}[t_s]) := (\boldsymbol{\pi}[t_{s-1}],\boldsymbol{p}[t_{s-1}])$ to constraints in S-SP2 (\ref{op:4})
\REPEAT
\STATE Solve (\ref{op:4}) to obtain $(\boldsymbol{\pi}^{*}[t_s],\boldsymbol{p}^{*}[t_s])$ and $\Xi^{*}[t_s]$;
\STATE Update\ \ $(\boldsymbol{\pi}^{(j)}[t_s],\boldsymbol{p}^{(j)}[t_s]) := (\boldsymbol{\pi}^{*}[t_s],\boldsymbol{p}^{*}[t_s])$ and $\Xi^{(j)}[t_s]:=\Xi^{*}[t_s]$;
\STATE Set $j:=j+1$;
\UNTIL Convergence or $|\Xi^{(j)}[t_s]-\Xi^{(j-1)}[t_s]|\leq \epsilon$ \{/*\textit{Satisfying a given accuracy level*}/\}\\
\STATE Recover an exact binary by computing $\boldsymbol{\pi}^{*}[t_s]=\lfloor \boldsymbol{\pi}^{(j)}[t_s] +0.5\rfloor$ and repeat step 1 to 5 for given $\boldsymbol{\pi}^{*}[t_s]$;
\STATE{\textbf{Output:}} $(\boldsymbol{\pi}^{*}[t_s],\boldsymbol{p}^{*}[t_s])$.
\end{algorithmic}
\end{algorithm}

\textbf{Penalty function:} We bring forward the following penalty function to accelerate the convergence of the proposed iterative algorithm that will be detailed shortly \[\mathcal{P}(\boldsymbol{\pi})=\sum_{t_s,f_i,m,u}\big[(\pi_{m,u,f_i}^{\sfem}[t_s])^2+(\pi_{m,u,f_i}^{\sfur}[t_s])^2 - \pi_{m,u,f_i}^{\sfem}[t_s]-\pi_{m,u,f_i}^{\sfur}[t_s] \big]\]
which is convex in $\boldsymbol{\pi}[t_s]$. It is clear that $\mathcal{P}(\boldsymbol{\pi})\leq 0$ for any $\pi_{m,u,f_i}^{\mathsf{x}}[t_s]\in[0,1]$, which is useful to penalize the relaxed variables to obtain near-precise binary solutions at optimum (\textit{i.e.} satisfying (\ref{sp:2b})). By incorporating 
$\mathcal{P}(\boldsymbol{\pi})$ into the objective function of (\ref{sp:2b}), the parameterized relaxed problem is expressed as:

\begin{subequations} \label{op:3}
\begin{align}
 \text{S-SP1}: \max_{\boldsymbol{\pi},\boldsymbol{p}}&\quad  \mathcal{R}^{\sfem} + \omega \mathcal{P}(\boldsymbol{\pi})  \\ 
 \text{s.t.} \quad & \quad \boldsymbol{\pi}[t_s]\in \tilde{\Pi}[t_s], \quad \forall t_s, \forall u \in \mathscr{U} \label{sp:3b}\\
  & (\ref{sp:2c})- (\ref{sp:2h})
\end{align}
\end{subequations}
 where $\tilde{\Pi}[t_s] \triangleq \{\pi_{m,u,f_i}^{\sfem}[t_s], \pi_{m,u,f_i}^{\sfur}[t_s] \in [0,1]|\sum_{u}[\pi_{m,u,f_i}^{\sfem}[t_s]+\pi_{m,u,f_i}^{\sfur}[t_s]]\leq 1$\} and $\omega>0$ denotes a determined penalty parameter.
\begin{proposition}
\textit{Problems (\ref{sp:2}) and (\ref{op:3}) share the same optimal solution, \textit{i.e.},} $(\boldsymbol{\pi}^{*},\boldsymbol{p}^{*})$, considering an suitable positive value of $\omega$.
\end{proposition}
The proof is directly followed \cite{ECheTWC14} by showing the fact that $\mathcal{P}(\boldsymbol{\pi})=0$ at optimum in maximizing of the objective function (\ref{op:3}). It implies that a constant $\omega$ always exists to guarantee that $\boldsymbol{\pi}$ are binary at optimum, and the relaxation is tight. Practically, it is acceptable if $\mathcal{P}(\boldsymbol{\pi})\leq \varepsilon$ for a tiny $\varepsilon$, which results in a nearly precise optimal solution.

In problem (\ref{op:3}), the objective function is non-concave due to $\mathcal{P}(\boldsymbol{\pi})$, while constraints (\ref{sp:2e}) is non-convex. Based on the SCA method, the first-order Taylor approximation is used to linearize the function $\mathcal{P}(\boldsymbol{\pi})$ at the $j$-th iteration as follows:

\begin{align}\label{eq12}
\nonumber \mathcal{P}^{(j)}(\boldsymbol{\pi})\triangleq& \sum_{m,u,f_i}\big[\pi_{m,u,f_i}^{\sfem}[t_s](2\pi_{m,u,f_i}^{\sfem,(j)}[t_s]-1) - (\pi_{m,u,f_i}^{\sfem,(j)}[t_s])^2 \nonumber\\
& + \pi_{m,u,f_i}^{\sfur}[t_s](2\pi_{m,u,f_i}^{\sfur,(j)}[t_s]-1) - (\pi_{m,u,f_i}^{\sfur,(j)}[t_s])^2 \big]
\end{align}
where $\mathcal{P}(\boldsymbol{\pi}) \geq \mathcal{P}^{(j)}(\boldsymbol{\pi})$ and $\mathcal{P}(\boldsymbol{\pi}^{(j)}) = \mathcal{P}^{(j)}(\boldsymbol{\pi}^{(j)})$.

To address constraint (\ref{sp:2e}), we indicate its LHS as $r_{m}(\boldsymbol{p}[t_s]) \triangleq \sum_u^{}\big[r_{m,u}^{\sfem}(\boldsymbol{p}^{\sfem}[t_s])+r_{m,u}^{\sfur}(\boldsymbol{p}^{\sfur}[t_s],$ $\boldsymbol{\pi}^{\sfur}[t_s])\big]$, which is concave in $\boldsymbol{p}[t_s]$. Thus, the function $r_{m}(\boldsymbol{p}[t_s])$ can be approximated at the feasible point $\boldsymbol{p}^{(j)}[t_s]$ as
\begin{align}\label{eq13}
	&r_{m}^{(j)}(\boldsymbol{p}[t_s])\triangleq  r_{m}(\boldsymbol{p}^{(j)}[t_s]) - \sum_{u,f_i}\beta_i \frac{\pi_{m,u,f_i}^{\sfur}[t_s] Q^{-1}(P_{e})}{\sqrt{ \delta_i \beta_i}} +\nonumber\\
    &\frac{\beta_i}{\ln2}\sum_{u,f_i,\mathsf{x}} (p^{\mathsf{x}}_{m,u,f_i}[t_s]-p^{\mathsf{x},(j)}_{m,u,f_i}[t_s]) \Big [ \frac{g_{m,u,f_i}[t_s]}{N_0+p^{\mathsf{x},(j)}_{m,u,f_i}g_{m,u,f_i}[t_s]} \Big].
\end{align}

The convex approximate program of (\ref{op:3}) solved at iteration $j$ is stated as follows, taking into account all the aforementioned approximations:
\begin{subequations} \label{op:4}
\begin{align}
\text{S-SP2}: \max_{\boldsymbol{\pi},\boldsymbol{p}}&\quad  \Xi^{(j)} \triangleq \mathcal{R}^{\sfem} + \omega \mathcal{P}^{(j)}(\boldsymbol{\pi})  \\ 
 \text{s.t.} \quad &  \eqref{sp:2c}, \eqref{sp:2d}, \eqref{sp:2f}- \eqref{sp:2h}, \eqref{sp:3b}\\
  &  r_{m}^{(j)}(\boldsymbol{p}[t_s]) \leq C_{m}^{\mathsf{FH}}, \forall m\in \mathscr{M}_n.
\end{align}
\end{subequations}
 
Algorithm \ref{alg2} provides a summary of the SCA-based iterative algorithm. Step 6 is used to recover an exact binary solution then Steps 1–5 are repeated to refine the final solution in order to ensure a feasible solution to the problem (\ref{op:3}).

\textit{Convergence and complexity analysis:} The development of the proposed iterative Algorithm \ref{alg2} is based on the SCA method \cite{Marks:78}. The approximations in \eqref{eq12} and \eqref{eq13} are satisfied the three key inner approximation properties given in \cite{Beck:JGO:10}, while other constraints are already linear and quadratic. In particular, the solution of \eqref{op:4} is always feasible to the parameterized relaxed
problem \eqref{op:3} but not vice versa. In addition, Algorithm \ref{alg2} generates a sequence of the improved solutions $\{\boldsymbol{\pi}^{(j)},\boldsymbol{p}^{(j)}\}$ in the sense that $\Xi^{(j+1)}\geq \Xi^{(j)}, \forall j$.  By \cite[Theorem 1]{Marks:78}, if the number of iterations is sufficiently large, the sequence $\{\boldsymbol{\pi}^{(j)},\boldsymbol{p}^{(j)}\}$ converges to at least a local optimal solution of \eqref{op:3}, satisfying the Karush-Kuhn-Tucker (KKT) conditions \cite[Theorem 1]{Marks:78}. On the other hand, for each numerology $i$, the convex approximate program \eqref{op:4} has $2MUF_i$ scalar decision variables and $2MUF_i + 4M + 3U$ linear and quadratic constraints. As a result, the worst-case computation complexity  of Algorithm \ref{alg2} in each iteration is estimated as $\mathcal{O}\bigl(\sqrt{2MUF_i + 4M + 3U}(2MUF_i)^3 \bigr)$, following the interior-point method \cite[Chapter 6]{Ben:2001}.


\section{Performance Evaluations And Numerical Results} \label{section6}
\subsection{Simulation Setup and Parameters}
We consider a scenario where all users are uniformly distributed in a circular area with a radius of $500$ m, while the locations of RUs are fixed. One RU is located in the central area, serving three sectors, each of which includes one RU. The RU-user channels are generated as Rayleigh fading with the path-loss $\mathsf{PL}_{\mathrm{RU-USER}}=128.1+37.6\log_{10}(d/1000)$ dB. The penalty factor is set to decrease after each TTI as $\omega[t_s]= 20+ 10/(1+t_s)$ to guarantee the convergence of the short-term subproblem. To estimate the future traffic for the upcoming frames, an RNN model's parameters, which include $2$ fully connected hidden layers and $50$ LSTM units (neurons), are trained. The operators can configure these parameters based on the provided data and its periodicity. The RNN training is carried out over the traffic dataset of the cellular network following Poisson distribution with the mean arrival rate of $20$ and $2.5$ for eMBB and uRLLC traffics, respectively. The mean arrival rate is a configurable parameter of the simulator. Incoming traffics packets are sorted in a first-come-first-serve buffer. The dataset contains network measurement in terms of arrival rate collected from $M$ RUs, over a horizon of $T = 10000$ traffic observations over a duration of $100$ seconds. The open-source, high-level TensorFlow version $1.13.1$ application programming interface, Keras, is used to implement the RNN model. All experiments are done on a Dell desktop computer with an Intel R CPU @ $3.0$ GHz. Simulation parameters including the LSTM model are summarized in Table \ref{tab1}. 

We put into practice the following five benchmark schemes for performance comparison:
\begin{enumerate}
\item \textit{Fixed numerology (FIX-NUM)}: In this scheme, the TTI is considered the same for both services as the LTE standard (\textit{i.e.} $0.5$ ms) with the SCS of $180$ kHz. The resource allocation, flow-split decision and dynamic BW-split for both traffics follow  Algorithm \ref{alg1} with some slight modifications. 

\item \textit{Equal Flow-Split Distribution (EFSD)}: In order to show the importance of optimizing the flow-split distribution per frame, this scheme considers the equal flow-split for each traffic to RUs, \textit{i.e.} $\varphi_{m,u}= \frac{1}{M},\ \forall u \in \mathscr{U}$.  

\item \textit{Equal Power Allocation (EPA)}: The RBs' allocation $\boldsymbol{\pi}$ is optimized by Algorithm \ref{alg1} for an equal power allocated to all users and subcarriers.

\item \textit{Single Connectivity with uRLLC Priority (SCUP)}: To reveal the performance improvement of MC in heterogeneous wireless networks, especially for eMBB throughput, this scheme provides the single connectivity (SC) scheme with uRLLC Priority. Due to the stringent requirement of latency, uRLLC will be predominantly guaranteed, and then the remaining resources are occupied by eMBB users. In this regard, this scheme considers $M$ RUs with disjoint dedicated users.

\item \textit{Proposed Problem in Presence of Known Traffic Demand (PKTD)}: This scheme investigates the performance of both traffics in the presence of known traffic demand $\lambda$. In practice, the obtained results of this scheme in the presence of unknown traffic demands show the accuracy of the LSTM model of the proposed method. 
\end{enumerate}

\begin{table}[!t]
\caption{Simulation Parameters}\vspace{-15pt}
\begin{center}
\begin{tabular}{|l|l|l|l|}
\hline
\textbf{Parameter} & \textbf{Value} &\textbf{Parameter} & \textbf{Value}\\
\hline\hline
 No. of RUs & 4 & Predetermined uRLLC latency ($D_{ur}$)& 0.5 ms\\
 \hline
 No. of eMBB users  & 12 & Predetermined eMBB throughput ($R_{th}$) & 1 Mbps\\
 \hline
 No. of uRLLC users  & 8 & Maximum FH capacity ($C^{FH}$)& 1 Gbps\\
\hline
 BW of RU & 20 MHz & Maximum MH capacity ($C^{MH}$) & 50 Gbps\\
 \hline
 Error probability ($P_e$)& $10^{-3}$ & Maximum RU's queue-length ($Q^{max}$) & 10 KB\\
 \hline
 Power of RU& 46 dBm & No. of LSTM layer& 2\\
 \hline
Noise power ($N_0$)& -110 dBm & No. of LSTM unit & 50\\
\hline
uRLLC packet size ($Z^{ur}$)& 1 KB & No. of epoch & 50\\
\hline
eMBB packet size ($Z^{em}$)& 125 KB & Activation function & \textit{tanh}\\
\hline
Length of time-frame& 10 ms & Optimizer & \textit{adam}\\
\hline
\end{tabular}
\label{tab1}
\end{center}
\end{table}

\begin{figure}
\centering
    \includegraphics[width=0.75\textwidth,trim=1 1 1 1,clip=true]{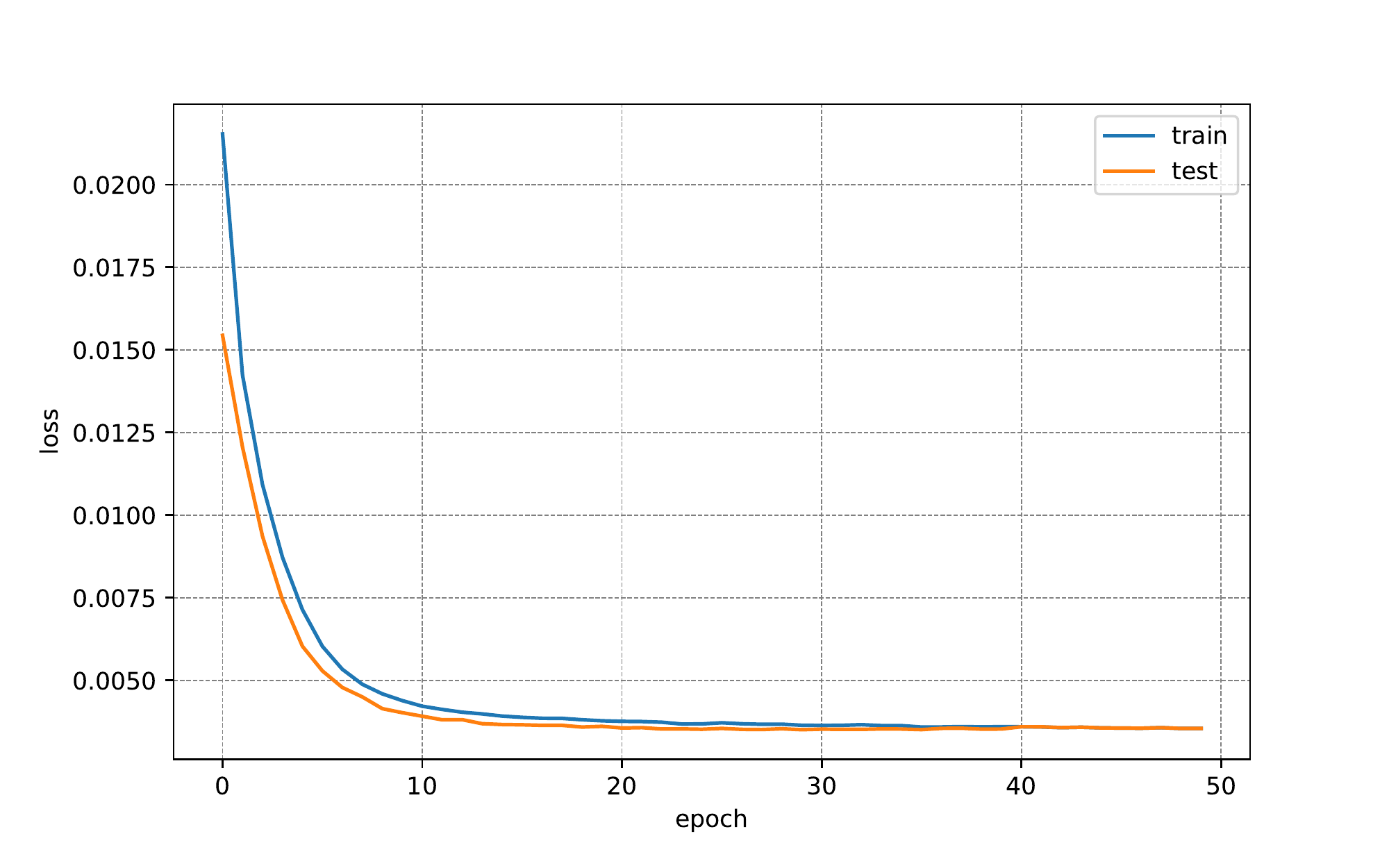}
    \caption{Training and validation loss for the LSTM RNN model.}
    \label{fig6}
\end{figure}

\subsection{Numerical Results and Discussions}

First, in order to investigate the LSTM's convergence, we monitor the value of the loss function as MSE and keep the training process until the training loss is typically identical to the validation loss after a specific number of epochs. Since the mean arrival rates of both traffics are not in the same range, we normalize traffic demands in the pre-processing phase through the MinMaxScaler normalization method from Sklearn. We then divide  data into two sets, which are $80\%$ for training and $20\%$ for validation. Fig. \ref{fig6} plots the training and validation losses for the LSTM model with the most suitable turning hyperparameters, which converge after 50 epochs. It should be mentioned that setting the desirable number of epochs prevents model overfitting. From Table \ref{tab2}, we find that the activation function of \textit{tanh} works better than \textit{relu} and \textit{sigmoid}. In the same condition, increasing the number of LSTM layers and decreasing the number of units per layer do not help reduce the MSE value. Based on the search result, the \textit{adam} optimizer converges faster than others, whereas it takes less time for the model's training. In our case, the dropout value is $0.01$ for both hidden layers. As a result, Table \ref{tab2} shows the search parameters to find the best parameters for the final LSTM-RNN model.

\begin{figure}
\centering
    \includegraphics[width=.95\textwidth,clip=true]{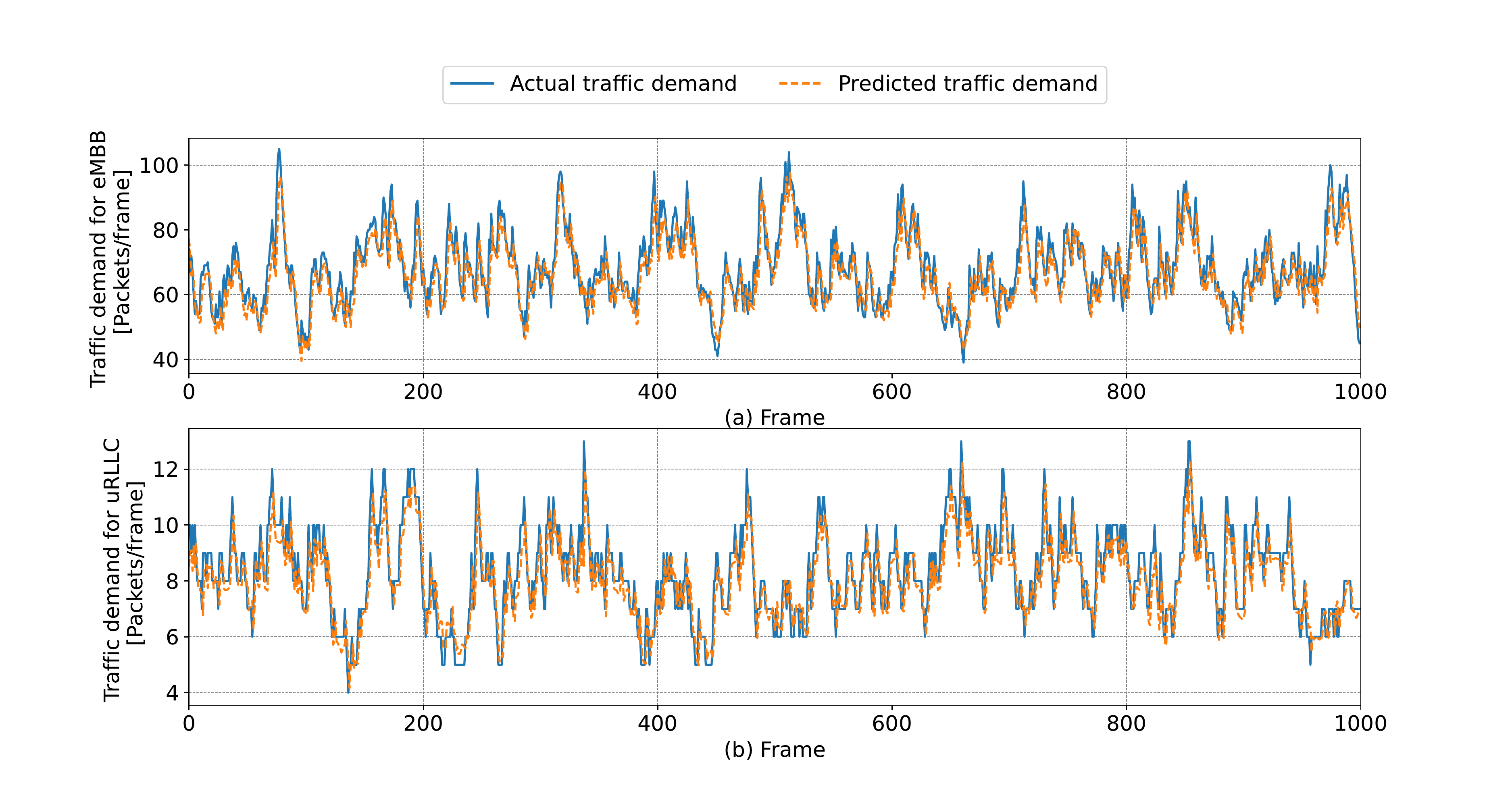}
    \caption{Traffic demand prediction in ORAN.}
    \label{fig7}
\end{figure}

The effectiveness of the LSTM RNN model in both traffic demands is represented in Fig. \ref{fig7} to illustrate the performance of the ML model prediction. The actual and predicted values for one of the eMBB and uRLLC traffic demands in the proposed system model are shown in Fig. \ref{fig7} (a) and Fig. \ref{fig7} (b), respectively. As it is clear from these figures, the trained LSTM-RNN model performs outstandingly in capturing the dynamic traffic demand of services over time. The difference between predicted and actual traffic demands is entirely small. The MSE value has been calculated as a performance measurement to validate the accuracy of the implemented LSTM model. For instance, the measured MSE values of the selected eMBB users in Fig. \ref{fig7} (a) and uRLLC users in Fig. \ref{fig7} (b) are $0.00315$ and $0.00323$, respectively.

\begin{table}[!t]
\caption{Hyperparameters for the different Performing LSTM Models}\vspace{-18pt}
\begin{center}
	\scalebox{0.88}{
\begin{tabular}{|p{3.6cm}|p{4.2cm}|p{2.5cm}|p{3cm}||p{2.5cm}|}
\hline
\textbf{No. of LSTM layers} & \textbf{No. of LSTM units in each layer} &\textbf{No. of epochs} & \textbf{Activation function}& \textbf{MSE}\\
\hline\hline
2& 20& 30& \textit{relu}& 0.00641\\
\hline
2& 50& 30& \textit{relu}& 0.00382\\
\hline
3& 50& 100& \textit{relu}& 0.00493\\
\hline
3& 50& 30& \textit{sigmoid}& 0.01281\\
\hline
2& 50& 30& \textit{sigmoid}& 0.00782\\
\hline
2& 50& 100& \textit{tanh}& 0.00421\\
\hline
3& 20& 30& \textit{tanh}& 0.00613\\
\hline
\textbf{2}& \textbf{50}& \textbf{50}& \textbf{\textit{tanh}}& \textbf{0.00331}\\
\hline
\end{tabular}
 }
\label{tab2}
\end{center}
\end{table}

\begin{figure}
\centering
    \includegraphics[width=0.80\textwidth,trim=1 1 1 1,clip=true]{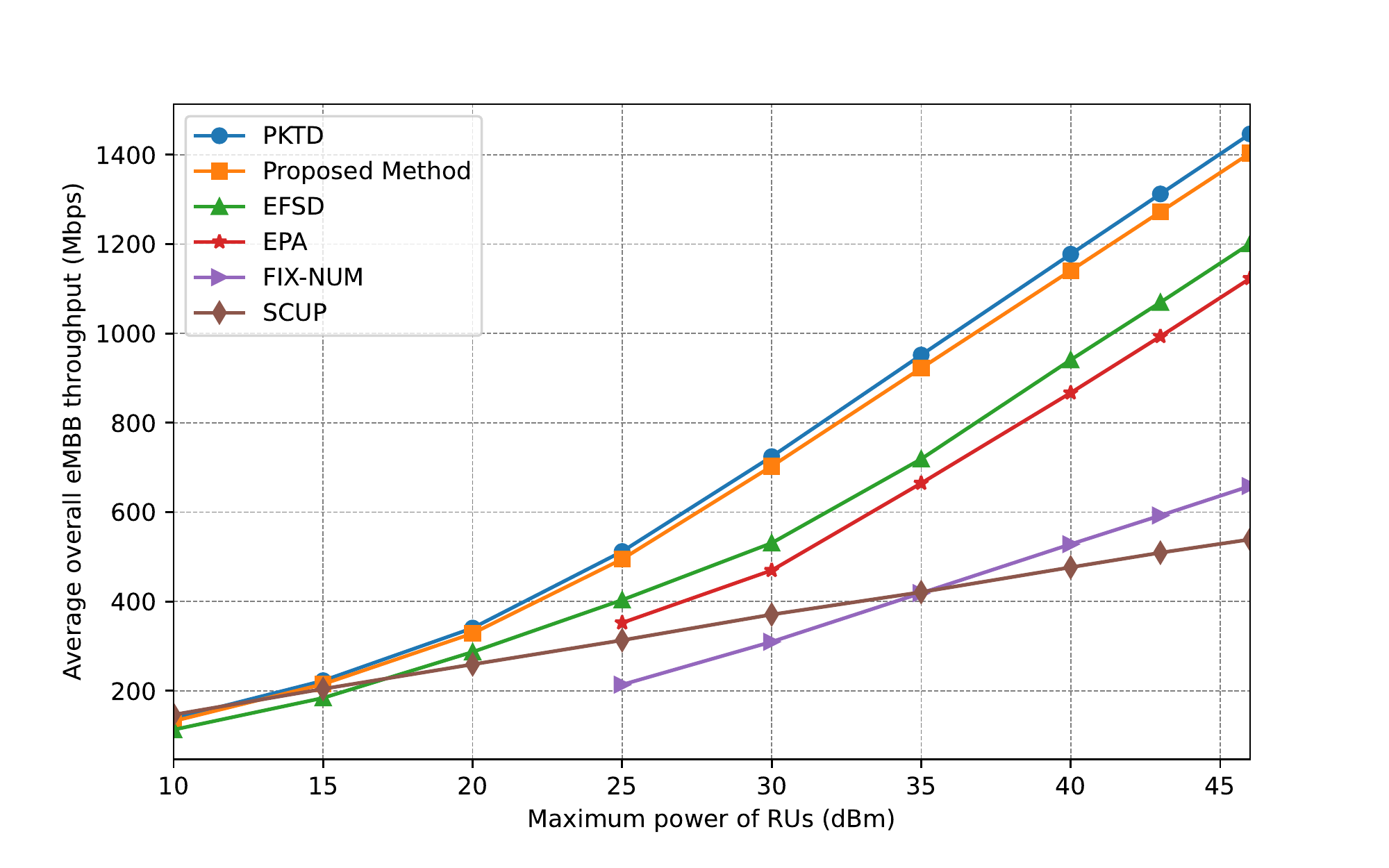}
    \caption{Average overall eMBB throughput versus $P^{\max}$.}
    \label{fig8}
\end{figure}

To evaluate the eMBB throughput with different resource allocation schemes, Fig. \ref{fig8} illustrates the sum throughput of eMBB users over different maximum RUs' power budgets from $10$ to $46$ dBm. Unsupringly, the PKTD provides the best performance and acts as the upper bound of all strategies. It can be observed that the gap between our proposed framework and PKTD is less than $3\%$, which proves the efficiency of the LSTM RNN model in predicting the dynamic traffic demand over time. Whereas the proposed method provides the highest eMBB throughput compared to other benchmark schemes. Comparing FIX-NUM, SCUP, EFSD and EPA, the proposed method offers $130.89\%$, $92.32\%$, $33.92\%$ and $51.21\%$ gains at the typical power value of $P^{\max}= 30$ dBm, respectively. Furthermore, EPA and FIX-NUM work over $P^{\max} \geq 25$ dBm, while they are infeasible when the maximum RUs' power less than $25$ dBm. Hence, this phenomenon shows the advantage of our proposed method over these schemes, especially at a small $P^{\max}$. Besides, as we mentioned previously, the MC technique plays a vital role in enhancing the eMBB throughput. The gap between the JIFDR framework considering the MC technique and SCUP grows with increasing the maximum power budget of RUs, representing that when the $P^{\max}$ is small, most users only link to one RU. While both JIFDR and SCUP have almost the same value at $P^{\max}= 10$ dBm, by increasing $P^{\max}$, the proposed method significantly exceeds that of SCUP. 
\begin{figure}
\centering
    \includegraphics[width=0.80\textwidth,trim=1 1 1 1,clip=true]{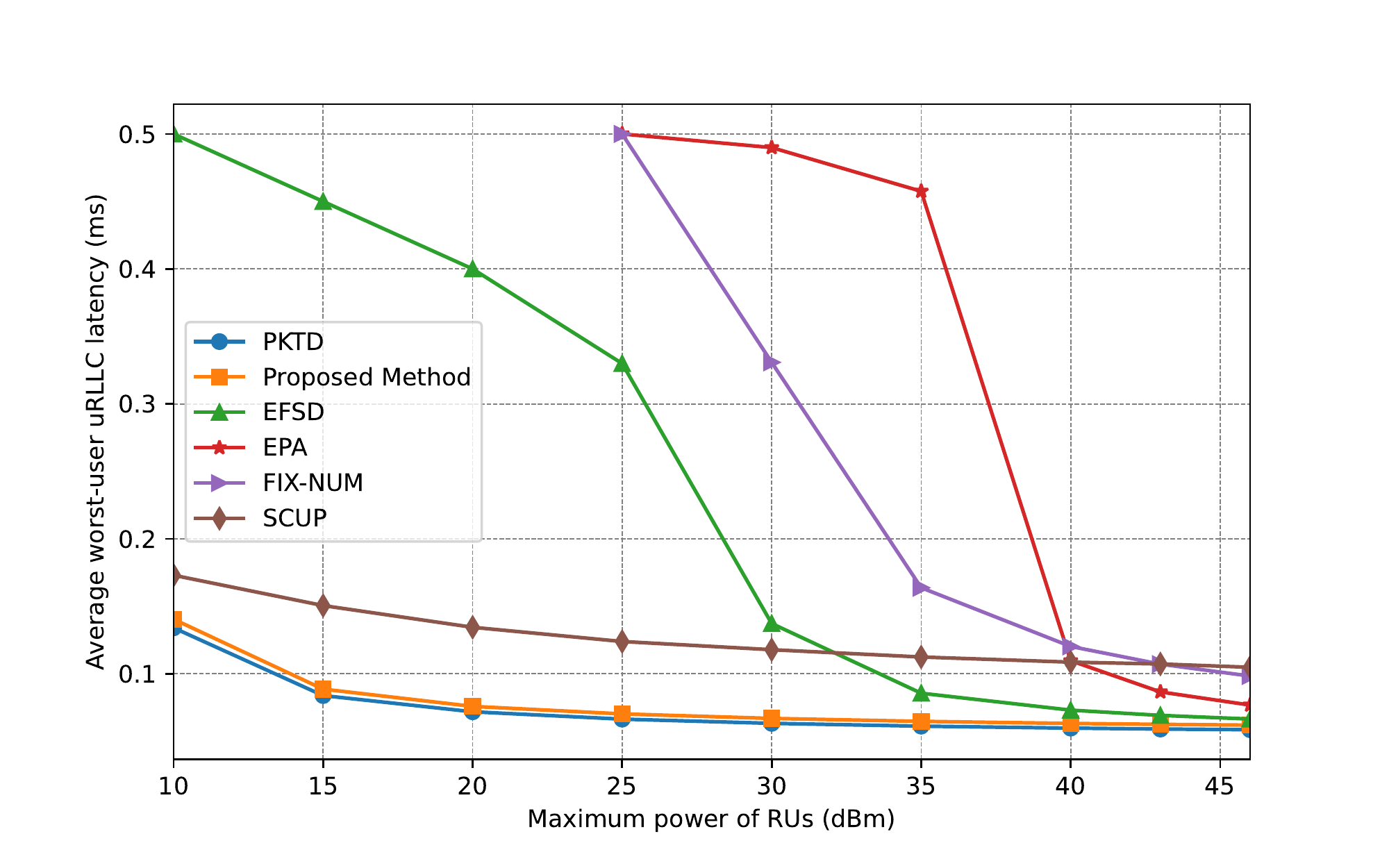}
    \caption{Average worst-user uRLLC latency versus $P^{\max}$.}
    \label{fig9}
\end{figure}

In order to show the performance of the proposed method on uRLLC latency, Fig. \ref{fig9} represents the worst-user uRLLC latency under different maximum power of RUs. Similar to the first optimization problem (P1), increasing the maximum power of RUs significantly affects the eMBB throughput improvement, resulting in an efficient reduction of uRLLC latency in the second optimization problem (P2). As we can see in Fig. \ref{fig9}, the uRLLC latency of the proposed method is almost equal to PKTD, which again confirms the accuracy of the LSTM RNN model in predicting the dynamic traffic demand. The performance gain in terms of latency of the proposed method is $84\%$ and $114.47\%$ Compared to SCUP and EFSD at $P^{\max}= 30$ dBm. According to the empty region of two benchmark schemes, FIX-NUM and EPA in the range $P^{\max} \leq 25$ dBm, results from Fig. \ref{fig9} show that these schemes are infeasible over the mentioned range of $P^{\max}$ while having a significant difference in uRLLC latency with the proposed method. Clearly, SCUP scheme in Fig. \ref{fig9} greatly outperforms SCUP scheme in Fig. \ref{fig8}.  On the one hand, the uRLLC and eMBB traffics are sliced in various virtual slices in SCUP, while the size of the uRLLC traffic packet is considerably smaller than eMBB packet size. Hence, the assigned slice to uRLLC could meet the uRLLC traffics' requirements alone without waiting in a queue. On the other hand, the SCUP scheme is not able to aggregate multiple links and allow users to connect to more than one RU to achieve the highest throughput.      

\begin{figure}
\centering
    \includegraphics[width=0.85\textwidth,trim=1 1 1 1,clip=true]{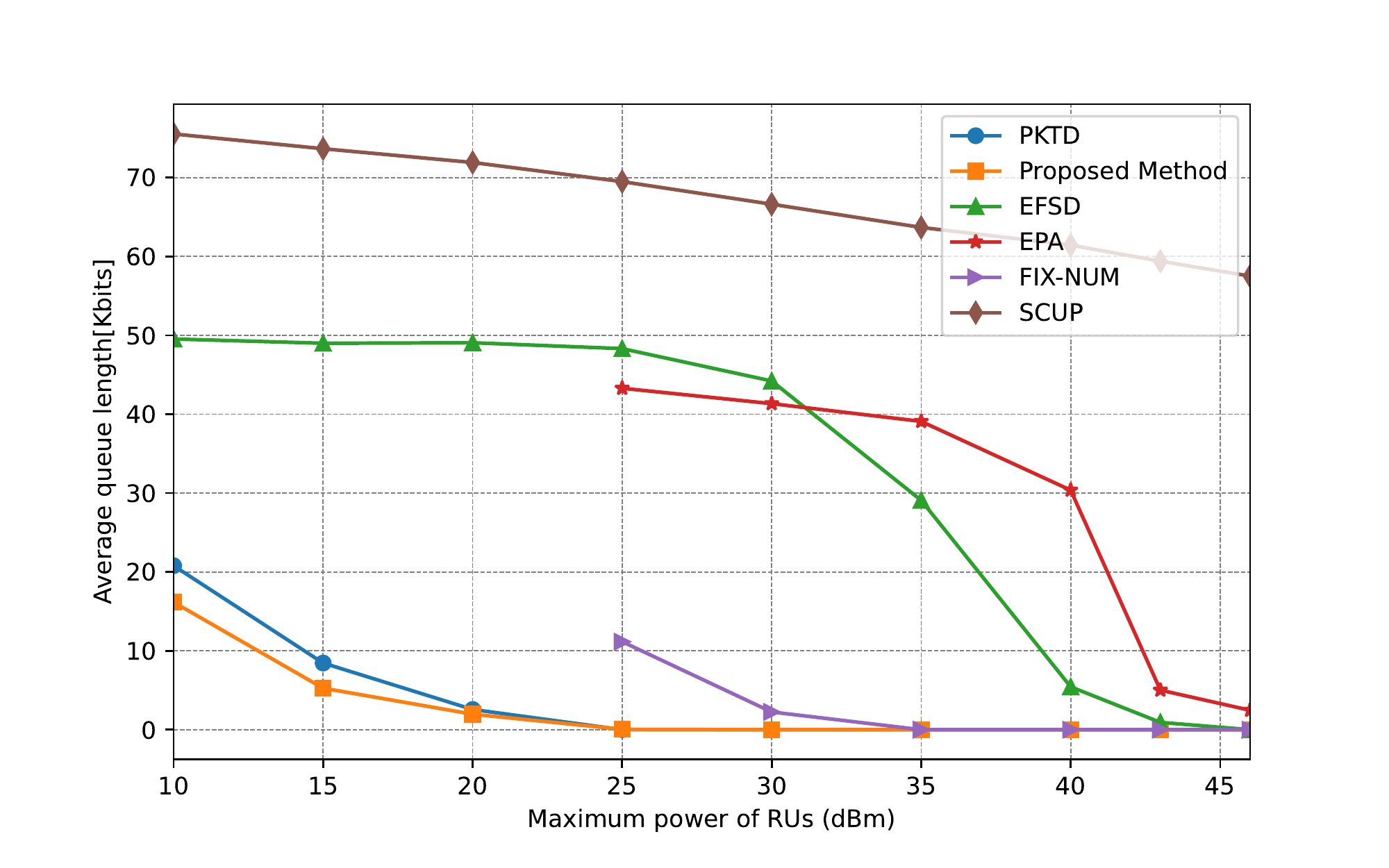}
    \caption{Average of queue lengths versus $P^{\max}$.}
    \label{fig10}
\end{figure}

Fig. \ref{fig10} depicts the average backlog in the queue under the maximum power budget of RUs with different benchmark schemes. As can be seen the higher the power budger $P^{\max}$, the lower the average queue-length. Similar to two previous figures, results from the proposed method and PKTD are very close to each other; meanwhile, both converge very fast to zero when $P^{\max} \geq 25$ dBm. As expected, the SCUP scheme yields the worst result, whereas the proposed method yields the best one. Two FIX-NUM and EPA schemes are infeasible when $P^{\max} < 25$ dBm. Clearly, the FIX-NUM benchmark scheme performs in a better way rather than EFSD, EPA and SCUP schemes for $P^{\max} \geq 25$ dBm. On the other hand, during the joint scheduling of uRLLC and eMBB traffics, we have numerically observed that  uRLLC users always prefer to have only one link in various system setups. This issue indicates that a single connection is generally the best option for traffic with small data packet size. In contrast, the MC technique is typically a nice option for traffic with high data packet size, \textit{i.e.} eMBB.

\begin{figure}
\centering
    \includegraphics[width=0.80\textwidth,trim=1 1 1 1,clip=true]{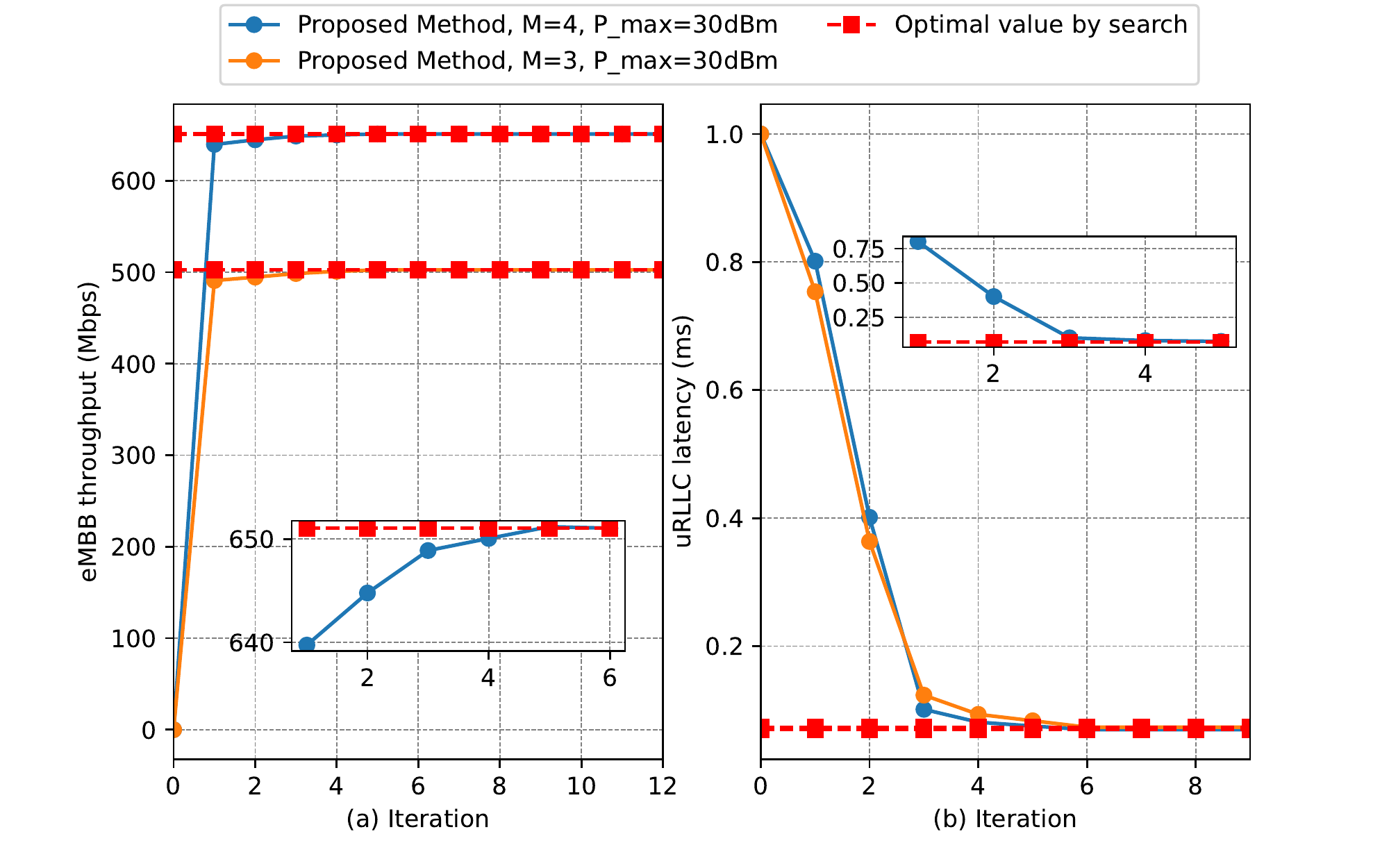}
    \caption{Convergence behaviour of the proposed Algorithm \ref{alg2}.}
    \label{fig11}
\end{figure}

Finally, we examine the convergence behavior of the proposed Algorithm \ref{alg1}, comparing the optimal value through the exhaustive search for $P^{\max}= 30$ dBm under the different number of RUs in Fig. \ref{fig11}. It is shown that the proposed algorithm for both problems (P1) and (P2) converges quickly, taking less than $10$ iterations to reach the optimal value within an increment, which is smaller than a given threshold $\epsilon = 10^{-4}$. As expected, based on Fig. \ref{fig11}(a) and Fig. \ref{fig11}(b), as the number of RUs increases in such a network, the eMBB throughput increases, but it does not affect the uRLLC latency remarkably. As we mentioned before that  uRLLC users frequently tend to link to only one RU because of their small packet size. There is almost the same convergence speed for both cases with 3 RUs and 4 RUs. Nevertheless, the case with $4$ RUs case needs a few more time for CVXPY to solve the MINCP in each step due to additional optimization variables.

\section{Conclusion} \label{section7}
In this work, we have developed a novel intelligent TS framework in the presence of unknown dynamic traffics to meet the competing demands of uRLLC and eMBB services in beyond 5G networks based on dynamic MC. To achieve the maximum throughput for eMBB traffic while guaranteeing the minimum uRLLC latency requirement, and vice versa, we have proposed a joint intelligent traffic prediction, flow-split distribution, dynamic RAN slicing and radio resource management scheme to schedule joint RBs and transmission power with mixed numerologies based on standardization in 5G NR. We have carried out a thorough analysis of E2E uRLLC latency. Due to the execution of the proposed problems in two different timescales, we have divided them into two long-term and short-term subproblems. To solve them, the LSTM method and  SCA-based iterative algorithm have been developed to solve the formulated subproblems effectively. Thanks to LSTM, which predicts future traffic with high accuracy, the proposed method based on MC and mixed numerologies greatly improves resource utilization by adapting to dynamic traffic demands compared to benchmark schemes.

\begingroup
\bibliography{ref1}

\begin{thebibliography}{10}

\bibitem{popovski20185g}
P.~Popovski {\em et~al.}, ``{5G wireless network slicing for eMBB, URLLC, and
  mMTC: A communication-theoretic view},'' {\em IEEE Access}, vol.~6,
  pp.~55765--55779, 2018.

\bibitem{gavrilovska2020cloud}
L.~Gavrilovska, V.~Rakovic, and D.~Denkovski, ``From cloud {RAN} to open
  {RAN},'' {\em Wirel. Pers. Commun.}, vol.~113, no.~3, pp.~1523--1539, 2020.

\bibitem{wang2020artificial}
C.-X. Wang, M.~Di~Renzo, S.~Stanczak, S.~Wang, and E.~G. Larsson, ``Artificial
  intelligence enabled wireless networking for {5G} and beyond: Recent advances
  and future challenges,'' {\em IEEE Wireless Communications}, vol.~27, no.~1,
  pp.~16--23, 2020.

\bibitem{niknam2020federated}
S.~Niknam, H.~S. Dhillon, and J.~H. Reed, ``Federated learning for wireless
  communications: Motivation, opportunities, and challenges,'' {\em IEEE
  Communications Magazine}, vol.~58, no.~6, pp.~46--51, 2020.

\bibitem{dryjanski2016unified}
M.~Dryjanski and M.~Szydelko, ``A unified traffic steering framework for {LTE}
  radio access network coordination,'' {\em IEEE Commun. Mag.}, vol.~54, no.~7,
  pp.~84--92, 2016.

\bibitem{vassilaras2017algorithmic}
S.~Vassilaras {\em et~al.}, ``The algorithmic aspects of network slicing,''
  {\em IEEE Commun. Mag.}, vol.~55, no.~8, pp.~112--119, 2017.

\bibitem{suer2019multi}
M.-T. Suer, C.~Thein, H.~Tchouankem, and L.~Wolf, ``Multi-connectivity as an
  enabler for reliable low latency communications—an overview,'' {\em IEEE
  Communications Surveys \& Tutorials}, vol.~22, no.~1, pp.~156--169, 2019.

\bibitem{arslan2018flexible}
H.~Arslan {\em et~al.}, ``Flexible multi-numerology systems for {5G} new
  radio,'' 2018.

\bibitem{yu2019review}
Y.~Yu, X.~Si, C.~Hu, and J.~Zhang, ``A review of recurrent neural networks:
  {LSTM} cells and network architectures,'' {\em Neural computation}, vol.~31,
  no.~7, pp.~1235--1270, 2019.

\bibitem{kamel2014lte}
M.~I. Kamel, L.~B. Le, and A.~Girard, ``{LTE} wireless network virtualization:
  Dynamic slicing via flexible scheduling,'' in {\em 2014 IEEE 80th Vehicular
  Technology Conference (VTC2014-Fall)}, pp.~1--5, IEEE, 2014.

\bibitem{pocovi2018joint}
G.~Pocovi, K.~I. Pedersen, and P.~Mogensen, ``Joint link adaptation and
  scheduling for {5G} ultra-reliable low-latency communications,'' {\em Ieee
  Access}, vol.~6, pp.~28912--28922, 2018.

\bibitem{karimi2019efficient}
A.~Karimi, K.~I. Pedersen, N.~H. Mahmood, G.~Pocovi, and P.~Mogensen,
  ``Efficient low complexity packet scheduling algorithm for mixed {URLLC and
  eMBB traffic in 5G},'' in {\em 2019 IEEE 89th Vehicular Technology Conference
  (VTC2019-Spring)}, pp.~1--6, IEEE, 2019.

\bibitem{wu2017signal}
Z.~Wu, F.~Zhao, and X.~Liu, ``Signal space diversity aided dynamic multiplexing
  for {eMBB and URLLC} traffics,'' in {\em 2017 3rd IEEE International
  Conference on Computer and Communications (ICCC)}, pp.~1396--1400, IEEE,
  2017.

\bibitem{anand2020joint}
A.~Anand, G.~De~Veciana, and S.~Shakkottai, ``Joint scheduling of {URLLC and
  eMBB traffic in 5G wireless networks},'' {\em IEEE/ACM Transactions on
  Networking}, vol.~28, no.~2, pp.~477--490, 2020.

\bibitem{zhang2016beyond}
N.~Zhang, S.~Zhang, S.~Wu, J.~Ren, J.~W. Mark, and X.~Shen, ``Beyond
  coexistence: Traffic steering in {LTE} networks with unlicensed bands,'' {\em
  IEEE Wireless Commun.}, vol.~23, no.~6, pp.~40--46, 2016.

\bibitem{korrai2020ran}
P.~Korrai, E.~Lagunas, S.~K. Sharma, S.~Chatzinotas, A.~Bandi, and
  B.~Ottersten, ``A {RAN} resource slicing mechanism for multiplexing of {eMBB
  and URLLC} services in {OFDMA based 5G wireless networks},'' {\em IEEE
  Access}, vol.~8, pp.~45674--45688, 2020.

\bibitem{zhang2020dynamic}
K.~Zhang, X.~Xu, J.~Zhang, B.~Zhang, X.~Tao, and Y.~Zhang, ``{Dynamic
  multiconnectivity based joint scheduling of eMBB and uRLLC in 5G networks},''
  {\em IEEE Systems Journal}, vol.~15, no.~1, pp.~1333--1343, 2020.

\bibitem{prasad2016enabling}
A.~Prasad, F.~S. Moya, M.~Ericson, R.~Fantini, and O.~Bulakci, ``Enabling {RAN}
  moderation and dynamic traffic steering in {5G},'' in {\em IEEE 84th Veh.
  Tech. Conf. (VTC-Fall)}, pp.~1--6, 2016.

\bibitem{you2018resource}
L.~You, Q.~Liao, N.~Pappas, and D.~Yuan, ``Resource optimization with flexible
  numerology and frame structure for heterogeneous services,'' {\em IEEE
  Communications Letters}, vol.~22, no.~12, pp.~2579--2582, 2018.

\bibitem{nguyen2019wireless}
T.~T. Nguyen, V.~N. Ha, and L.~B. Le, ``Wireless scheduling for heterogeneous
  services with mixed numerology in {5G} wireless networks,'' {\em IEEE
  Communications Letters}, vol.~24, no.~2, pp.~410--413, 2019.

\bibitem{korrai2020joint}
P.~K. Korrai, E.~Lagunas, A.~Bandi, S.~K. Sharma, and S.~Chatzinotas, ``Joint
  power and resource block allocation for mixed-numerology-based {5G downlink
  under imperfect CSI},'' {\em IEEE Open Journal of the Communications
  Society}, vol.~1, pp.~1583--1601, 2020.

\bibitem{niknam2020intelligent}
S.~Niknam {\em et~al.}, ``Intelligent {O-RAN} for beyond {5G and 6G} wireless
  networks,'' 2020. [Online]:.
\newblock \url{https://arxiv.org/abs/2005.08374}.

\bibitem{bonati2021intelligence}
L.~Bonati {\em et~al.}, ``{Intelligence and learning in O-RAN for data-driven
  NextG cellular networks},'' {\em IEEE Commun. Mag.}, vol.~59, no.~10,
  pp.~21--27, 2021.

\bibitem{kavehmadavani2022traffic}
F.~Kavehmadavani, V.-D. Nguyen, T.~X. Vu, and S.~Chatzinotas, ``{Traffic
  Steering for eMBB and uRLLC Coexistence in Open Radio Access Networks},'' in
  {\em 2022 IEEE International Conference on Communications Workshops (ICC
  Workshops)}, pp.~242--247, IEEE, 2022.

\bibitem{oran2018}
ORAN\hspace{2pt}Alliance, ``{O-RAN: Towards an open and smart RAN}.''
  \url{https://www.o-ran.org/resources}, 2018.

\bibitem{kihero2019inter}
A.~B. Kihero, M.~S.~J. Solaija, and H.~Arslan, ``Inter-numerology interference
  for beyond {5G},'' {\em IEEE Access}, vol.~7, pp.~146512--146523, 2019.

\bibitem{polyanskiy2010channel}
Y.~Polyanskiy, H.~V. Poor, and S.~Verd{\'u}, ``Channel coding rate in the
  finite blocklength regime,'' {\em IEEE Transactions on Information Theory},
  vol.~56, no.~5, pp.~2307--2359, 2010.

\bibitem{schiessl2015delay}
S.~Schiessl {\em et~al.}, ``Delay analysis for wireless fading channels with
  finite blocklength channel coding,'' in {\em Proc. 18th ACM Inter. Conf.
  Model. Anal. and Simul. Wire. and Mob. Sys.}, pp.~13--22, 2015.

\bibitem{burke1956output}
P.~J. Burke, ``The output of a queuing system,'' {\em Oper. Res.}, vol.~4,
  no.~6, pp.~699--704, 1956.

\bibitem{alliance2019ran}
O.~Alliance, ``{O-RAN Working Group 2 AI/ML Workflow Description and
  Requirements},'' {\em ORAN-WG2. AIML. v01}, vol.~1, 2019.

\bibitem{ECheTWC14}
E.~Che {\em et~al.}, ``Joint optimization of cooperative beamforming and relay
  assignment in multi-user wireless relay networks,'' {\em IEEE Trans. Wireless
  Commun.}, vol.~13, no.~10, p.~5481–5495, 2014.

\bibitem{Marks:78}
B.~R. Marks and G.~P. Wright, ``A general inner approximation algorithm for
  nonconvex mathematical programs,'' {\em Operations Research}, vol.~26,
  pp.~681--683, July-Aug. 1978.

\bibitem{Beck:JGO:10}
A.~Beck, A.~Ben-Tal, and L.~Tetruashvili, ``A sequential parametric convex
  approximation method with applications to nonconvex truss topology design
  problems,'' {\em J. Global Optim.}, vol.~47, pp.~29--51, May 2010.

\bibitem{Ben:2001}
A.~Ben-Tal and A.~Nemirovski, {\em Lectures on Modern Convex Optimization.}
\newblock Philadelphia: MPS-SIAM Series on Optimi., SIAM, 2001.

\end{thebibliography}
\bibliographystyle{ieeetr}
\endgroup
\end{document}